\documentclass[12pt]{article}
 \usepackage{tcolorbox}

\usepackage{graphicx,verbatim,array,multicol, xcolor, lscape, mathrsfs, color,  algorithm, algorithmicx, algpseudocode}
\usepackage{psfrag, amsmath, amsfonts, epsfig, fancybox, setspace,soul,amsthm,natbib, subcaption}
\usepackage{pdflscape}
\usepackage{xr-hyper}
\usepackage[hidelinks]{hyperref}

\def\bS{{\bf S}}

\def\bSigma{\boldsymbol{\Sigma}}
\def\bPi{\boldsymbol{\Pi}}
\def\by{{\bf y}}
\def\supone{^{(1)}}
\def\supzero{^{(0)}}
\def\supg{^{(g)}}

\def\bbeta{\boldsymbol{\beta}}

\def\bSigma{\boldsymbol{\Sigma}}
\def\bPi{\boldsymbol{\Pi}}

\def\bS{{\bf S}}

\def\by{{\bf y}}
\def\supone{^{(1)}}
\def\supzero{^{(0)}}
\def\supg{^{(g)}}

\def\bK{{\bf K}}
\def\bM{{\bf M}}

\def\bbeta{\boldsymbol{\beta}}

\newtheorem{theorem}{Theorem}
\usepackage{afterpage}
\usepackage{ltablex} 

\newtheorem{assumption}{Assumption}

\begin{document}
\thispagestyle{empty}

\begin{center}
\LARGE{Resilience Measures for the Surrogate Paradox}\\

\vspace*{15mm}
\normalsize Emily Hsiao$^{1}$, Lu Tian$^{2}$, and Layla Parast$^{3}$ \\
\vspace*{10mm}
\normalsize $^{1}$Department of Statistics and Data Sciences, University of Texas at Austin, \\ \normalsize 105 E 24th St D9800, Austin, TX 78705 \\
\normalsize $^{2}$Department of Biomedical Data Science, Stanford University, \\ \normalsize 1265 Welch Road, Stanford, CA 94305\\
\normalsize $^{3}$Department of Statistics and Data Sciences, University of Texas at Austin, \\ \normalsize 105 E 24th St D9800, Austin, TX 78705, parast@austin.utexas.edu \\

\end{center}

\clearpage
\thispagestyle{empty}
\begin{abstract}
Surrogate markers are often used in clinical trials to evaluate treatment effects when primary outcomes are costly, invasive, or take a long time to observe. However, reliance on surrogates can lead to the ``surrogate paradox,” where a treatment appears beneficial based on the surrogate but is actually harmful with respect to the primary outcome. In this paper, we propose formal measures to assess resilience against the surrogate paradox. Our setting assumes an existing study in which the surrogate marker and primary outcome have been measured (Study A) and a new study (Study B) in which only the surrogate is measured. Rather than assuming transportability of the conditional mean functions across studies, we consider a class of functions for Study B that deviate from those in Study A. Using these, we estimate the distribution of potential treatment effects on the unmeasured primary outcome and define resilience measures including a resilience probability, resilience bound, and resilience set. Our approach complements traditional surrogate validation methods by quantifying the plausibility of the surrogate paradox under controlled deviations from what is known from Study A. We investigate the performance of our proposed measures via a simulation study and application to two distinct HIV clinical trials.
\end{abstract}


\clearpage
\setcounter{page}{1}
\section{Introduction}
\vspace*{-0.4cm}
\subsection{The problem with surrogate markers}
Surrogate markers are commonly used in clinical studies to replace a primary outcome in evaluating or comparing the effectiveness of a new treatment \citep{FDA_approved}. In the United States (US), drugs can be approved early based on demonstrated effectiveness on a surrogate marker through the controversial Accelerated Approval Program of the US Food and Drug Administration (FDA) \citep{FDA_program}. For example, Jardiance was originally approved by the FDA for people with Type 2 diabetes based on results showing that the drug had an effect on a surrogate marker, blood glucose level \citep{Lily2014}. Subsequent results two years later showed that the drug indeed reduced the risk of cardiovascular death in these patients \citep{FDA_drug}. More recently, the FDA approved the Alzheimer's drug aducanumab based on results showing that it reduced visible plaque in the brain. However, later results revealed that the drug had no impact on patient outcomes, and it was voluntarily withdrawn from the market by the manufacturer \citep{dunn2021approval,alexander2021revisiting, biogen}.

In settings where the surrogate marker can be measured earlier than the primary outcome, the potential benefits of using the surrogate include more timely treatment approval (or treatment failure) decisions and reduced patient follow-up. For example, in the Jardiance example above, considering blood glucose level as a surrogate for cardiovascular death led to earlier approval and availability of the drug. In contrast, in settings where the surrogate marker would be measured at the same time as the primary outcome, but is less costly or invasive, the potential benefits of using the surrogate include reduced costs and reduced patient burden. For example,  in studies evaluating treatments for individuals with non-alcoholic fatty liver disease (NAFLD), using serum biomarkers like alanine aminotransferase as a surrogate for the NAFLD activity score (which is typically measured through liver biopsy) can eliminate the need for a biopsy. This is particularly advantageous in pediatric NAFLD studies, where avoiding a liver biopsy is especially appealing \citep{lavine2011effect,sanyal2023noninvasive}. 

With these potential benefits come great potential risks. Two main types of errors can occur. First, one may conclude that a treatment is ineffective based on the surrogate marker results when it in fact does have a positive effect on the primary outcome. This is important but generally considered to be of less concern. Second, one may conclude that a treatment is effective based on surrogate marker results when in fact it has no effect on the primary outcome (as in the aducanumab example) or worse, has a harmful effect on the primary outcome. It is this latter situation that is considered the most unacceptable of risks and is referred to as the ``surrogate paradox" \citep{chen2007criteria, vanderweele2013surrogate}. It is of utmost importance to ensure that the surrogate paradox does not occur. We describe the surrogate paradox problem as follows. Assume we have a surrogate marker that has been deemed valid based on data and analysis of a prior study, referred to as Study A. In Study A, both the surrogate and the primary outcome were measured. We have now conducted a new study, referred to as Study B, to evaluate the effect of the treatment and we have only measured the surrogate (not the primary outcome). There is evidence of a beneficial treatment effect on the surrogate marker in Study B. How can we guarantee, if at all, that there is not a harmful treatment effect on the primary outcome in Study B? That is, how can we guarantee that the surrogate paradox is not present in Study B? Our goal in this paper is to develop an empirical approach to answer this question. We refer to this concept as ``resilience against the surrogate paradox" as resilience refers to the ability to adapt or recover from difficult situations. We propose multiple ``resilience measures" and algorithms to estimate the proposed measures.

Our focus is not on methods to validate a surrogate in Study A, a topic on which much 
research has been done and for which we refer readers to \citet{elliott2023surrogate}. 
Throughout, we focus on developing statistical methods to address the surrogate paradox question described above with the understanding that the importance of clinical validity and discussion are paramount. We do not have, nor claim to have, the clinical knowledge to offer insight on assessing this question from a clinical perspective, but we argue that this perspective should be addressed prior to or in parallel with our proposed methods.

\subsection{A brief look at our contribution}
This paper addresses the problem of assessing the risk of a surrogate paradox when we have data from a previous study validating the surrogate and a subsequent study in which the surrogate, but not the primary outcome, is collected. Information about the relationship between the surrogate and the outcome from the previous study will be used to make a conclusion about the treatment effect on the primary outcome, using the surrogate only, in the subsequent study.  

More formally, let $G$ be a binary randomized treatment indicator, where $G = 1$ for the treatment group and $G = 0$ for the control group. Let $S$ denote the value of the surrogate and $Y$ denote the primary outcome of interest. Borrowing from the potential outcomes notation, we use $S_K\supg$ and $Y_K\supg$ to refer to the surrogate and primary outcome under treatment $g$ in Study $K$ for $g=0,1$ and $K=A,B$. Without loss of generality, we assume that higher values of $S$ and $Y$ are better. Data from that the previous study, which we refer to as Study A, include the surrogate measurements, primary outcome, and treatment group. That is, in Study A, we observe $\{S_{A0i}, Y_{A0i}\}$ for subjects $i = 1,...,n_{A0}$ for the control group and $\{S_{A1i}, Y_{A1i}\}$ for subjects $i = 1,...,n_{A1}$ for the treatment group. In the subsequent study, which we refer to as Study B, the primary outcome is not available. That is, we only observe $\{S_{B0i}\}$ for subjects $i = 1,...,n_{B0}$ in the control group and $\{S_{B1i}\}$ for subjects $i = 1,...,n_{B1}$ in the treatment group; no $Y$ values are observed.

The treatment effect on $S$ in Study B is denoted as $\Delta_{SB} = E(S_B\supone - S_B\supzero)$ and is both identifiable and estimable. The treatment effect on $Y$ in Study B is denoted as  $\Delta_B = E(Y_B\supone - Y_B\supzero)$ and is the ultimate quantity of interest, though it not directly estimable since $Y$ is not measured in Study B. Of course, the main motivation behind identifying and using a surrogate marker is so that we can make inference about $\Delta_B$ when only $S$ is available in Study B. It should come as no surprise that in order to achieve this goal, one must make certain assumptions about the transportability of surrogate information from Study A to Study B. For example, existing methods that offer procedures for robust inference about $\Delta_B$ generally assume transportability of, at a minimum, the conditional mean from Study A to Study B i.e., $\mu_{Ag}(s) = \mu_{Bg}(s)$, where $\mu_{Kg}(s) \equiv  E(Y_K^{(g)}|S_K^{(g)}=s)$ \citep{parast2019using,parast2024group}. Here, we instead ask, what if $\mu_{Bg}(s)$ in Study B is not the same as Study A, but is ``not too far away"? In addition, in this paper, we don't truthfully care about our ability to estimate $\Delta_B$; we only care about whether $\Delta_B$ is negative $(\Delta_B<0)$, as this would indicate the presence of the surrogate paradox.

To summarize our objective, we have no intention to examine the strength of the surrogate in Study A. Instead, we assume that there is sufficient evidence (using any chosen surrogate validation framework) that $S$ is a good surrogate marker in Study A such that the treatment effect on $S$ can be used to make inference about the treatment effect on $Y$ via $\mu_{Ag}(s), g=0, 1.$. One hopes that the relationship between $Y^{(g)}_B$ and $S^{(g)}_B$, quantified by $\mu_{Bg}(s),$ bears enough similarity to $\mu_{Ag}(s)$. Our goal is to quantify this similarity given the estimated $\mu_{Ag}(s), g=0, 1$ in Study A and the observed surrogate marker values in Study B. To this end, our proposed approach considers alternative $\mu_{Bg}$ functions for Study B within a general class of functions that are within a certain distance from the observed functions from Study A. We suggest that one may generate new potential $\mu_{Bg}$ functions within a particular class (described later), and use these functions to examine the resulting potential $\Delta_B$'s, inputting the observed values of $S$ which are available from Study B. Repeating this process, we obtain a ``distribution" of potential $\Delta_B$ values and use this distribution to define the \textit{resilience probability}. If the resilience probability is large, this suggests that the surrogate paradox is possible. If the resilience probability is small, this suggests that the surrogate paradox is perhaps unlikely. In addition, we use this ``distribution" to define a \textit{resilience bound} such that if the resilience bound is negative, this suggests that the surrogate paradox is possible; if it is positive, this suggests that the surrogate paradox is perhaps unlikely. Lastly, we propose a \textit{resilience set} which is the set of $\mu_{Bg}$ functions within the specified class where the resilience bound is greater than some desired threshold. Our proposed approach essentially asks, how extreme could $\Delta_B$ hypothetically be? And, more precisely, how different do the $\mu_g$ functions have to be in Study B (compared to Study A) to make $\Delta_B$ be negative? If only by considering $\mu_{Bg}$ functions that are very far from the observed $\mu_{Ag}$ in Study A do we obtain a high likelihood of a negative $\Delta_B$, then perhaps Study B is resilient to the surrogate paradox. In contrast, if generating $\mu_{Bg}$ functions only slightly different from the observed $\mu_{Ag}$ in Study A results in a high likelihood of a negative $\Delta_B$, then perhaps Study B is \textit{not} resilient to the surrogate paradox. This idea is similar to and inspired by existing work that uses simulations and formal bounds to assess robustness to unmeasured confounders in which one simulates or postulates characteristics of unmeasured confounders \citep{liu2013introduction,chernozhukov2022long,higashi2005quality,lin1998assessing}. Methods for robustness to unmeasured confounders asks, how extreme does an unmeasured confounder have to be to reverse my study's conclusion? If it takes a very extreme confounder to reverse the conclusion, one may decide that their conclusion is robust to unmeasured confounding. If it takes a very weak confounder to reverse the conclusion, one may decide that their conclusion is \textit{not} robust to unmeasured confounding.

Of course, the difficulty lies in deciding how we generate these potential functions. The majority of the details in this paper is devoted to various approaches to generate such functions and use them to estimate our proposed resilience measures. 

\subsection{Relationship to existing work}

Our problem of interest and proposed approach has connections to several related research areas including existing surrogate paradox assessment approaches, the surrogate threshold effect, domain adaptation, conformal inference, post-prediction inference, and adversarial machine learning. We briefly discuss each below.

\subsubsection{Sufficient conditions to protect against the surrogate paradox}
In the surrogate literature, much work has been done in identifying sufficient conditions to preclude the surrogate paradox. \citet{chen2007criteria} and \citet{ju2010criteria} have identified conditions to protect against the surrogate paradox if one assumes that the treatment has no direct effect on the primary outcome that does not pass through the surrogate (i.e., perfect surrogate). \citet{wang2002measure} have also proposed three sufficient but not necessary conditions to protect against the surrogate paradox and \citet{hsiao2025} has developed a series of statistical tests to empirically test these conditions using data from Study A, while \citet{guo2024quantifying} proposed a different framework slightly relaxing these assumptions. However, there are two difficulties with such methods. First of all, these assumptions are sufficient but not necessary, and thus, it is possible for the conditions to be violated without the presence of the surrogate paradox. Though work has been done on more specific bounds in the binary case in \citep{yin2020novel}, in the general case, there exists significant ambiguity in interpreting violations of the conditions. The second difficulty is that these assumptions may only ever be evaluated in Study A, as we do not collect the primary outcome in Study B. (If the primary outcome is collected in Study B, this discussion is moot because we can directly estimate $\Delta_B$). That is, even if we confirm that these conditions hold in Study A, the only way they would protect us from the surrogate paradox is if we make the strong assumption that the $\mu_{Kg}(s)$ functions remain exactly the same between studies, which is unlikely to reflect reality. 

\subsubsection{Surrogate paradox risk in a meta-analytic setting}

Prior work by \citet{elliott2015surrogacy} and \citet{shafie2023incorporating} has offered a method of evaluating the probability of a surrogate paradox in a meta-analytic setting by assuming the distribution of trial-level treatment effects are multivariate normally distributed. These methods are useful in that they directly target the quantifiable probability that, given a certain number of existing trials with the same $S$ and $Y$, a new trial where we only have $S$ will have the surrogate paradox. This approach is ideal in a setting where we have many existing trials and when the  underlying parametric assumptions are true, but is not applicable when we have only a single prior study (Study A), which is our case.

\subsubsection{Surrogate Threshold Effect}
The surrogate threshold effect (STE), introduced by \citet{burzykowski2006surrogate}, is the minimum treatment effect on the surrogate necessary to predict a meaningful effect on the primary outcome. When focused on the surrogate paradox, interest lies in the STE necessary to ensure a non-negative treatment effect on the primary outcome. The STE provides a practical decision rule: if a new treatment achieves at least the STE on the surrogate, it is likely to have a non-negative effect on the primary outcome. While informative, the STE has limited utility when only a single prior study is available because its estimation relies on the joint distribution of treatment effects across multiple studies to establish a robust threshold. 

\subsubsection{Domain adaptation}\label{domain_adaptation}
Recent extensive work in domain adaptation has addressed the challenge of distributional differences between source and target domains, which is clearly relevant to our setting \citep{shimodaira2000improving, farahani2021brief, kouw2019review}. Here, we can consider Study~A as the source domain and Study~B as the target, and our goal is to quantify how the treatment effect might differ between the two. Domain adaptation methods often focus on \emph{covariate shift}, where the marginal distribution of covariates (in our case, the surrogate $S$) differs between domains, while the conditional outcome distribution remains stable. Our setting, however, involves both potential changes in the \emph{conditional} distribution of the primary outcome given the surrogate, commonly referred to as \emph{concept shift}, and covariate shift. While robust estimation methods have been proposed for this problem, they often rely on the assumption that certain source-domain statistics remain unchanged in the target domain. This is not appropriate for our application, where we are explicitly interested in whether quantities like the average treatment effect on the primary outcome differ in the new study. 

\subsubsection{Conformal inference}
Our proposed approach shares some philosophical overlap with conformal inference but is fundamentally different. Conformal inference provides finite-sample, distribution-free predictive or confidence sets with guarantees under the assumption of exchangeability, and is primarily focused on predictive uncertainty for individual-level outcomes \citep{shafer2008tutorial}. In contrast, we focus on evaluating the plausibility of a harmful treatment effect on the primary outcome when only surrogate data are available in a new study, leveraging partial information transported from a previous study. While some conformal methods incorporate weights to address covariate shift between studies, their validity typically hinges on the assumption that the conditional distribution remains unchanged. By contrast, we explicitly allow the conditional distribution—in our case, the conditional mean of the primary outcome given the surrogate—to differ between Study A and Study B. Rather than assume invariance, we define a flexible approach to capture potential changes in this relationship, and use it to assess the likelihood of the surrogate paradox. Instead of targeting formal coverage guarantees, our approach defines a class of plausible conditional mean functions and simulates corresponding treatment effects to quantify resilience against the surrogate paradox.

\subsubsection{Prediction-Powered Inference}
Our proposed framework shares conceptual parallels with prediction-powered inference (PPI) \citep{angelopoulos2023prediction,xu2025unified}. PPI seeks to estimate a parameter $\theta$ from an unlabeled dataset $(\widetilde{X})_{i=1}^n$ by leveraging a labeled dataset $(X,Y)_{i=1}^N$ and an external prediction rule $f$ which is used to obtain predictions in the unlabeled dataset. PPI corrects for systematic bias in $f$ using a rectifier that explicitly measures prediction error on the labeled data, enabling valid inference for $\theta$ even when $f$ is imperfect. Both PPI and our proposed approach involve a prediction rule and separate labeled and unlabeled samples, but there are critical distinctions between our setting and PPI. First, the prediction function $\mu_{Kg}(s)$ in our framework is not externally supplied but estimated from the labeled data (Study A), and thus inherently tied to it. Second, PPI assumes the labeled and unlabeled data arise from identical distributions, whereas we explicitly relax this assumption and investigate sensitivity to distributional shifts. Finally, PPI requires some labeled observations from the target distribution to estimate rectification bias, whereas in our setting, no labels are observed in Study B, rendering direct application of PPI infeasible. 
\subsubsection{Adversarial Minimax Methods}
Our framework also bears a surface resemblance to adversarial minimax methods developed for treatment effect prediction \citep{chu2020treatment,du2021adversarial}. In these approaches, often inspired by adversarial machine learning, a counterfactual generator and a discriminator are trained in opposition: the generator aims to produce counterfactual outcomes indistinguishable from observed data, while the discriminator aims to detect them, yielding minimax optimization of predictive performance under worst-case conditions. In contrast, our goal is not to generate counterfactuals that mimic observed outcomes, nor to minimize prediction error under adversarial loss. Rather, assuming a randomized trial setting, we aim to explore a range of alternative conditional mean functions consistent with bounded deviations from the observed conditional mean functions. Our goal is to understand the potential behavior of $\Delta_B$ under distributional perturbations, not to achieve indistinguishability between Study A and Study B outcomes. Thus, while adversarial training methods and our framework both consider worst-case scenarios, the targets and motivations are fundamentally different.
\vspace*{-0.4cm}
\subsection{Outline of the Paper}
The remainder of the paper is structured as follows: Section \ref{proposed} formally states the problem, defines the resilience measures, and introduces various algorithms for estimation. Section \ref{inference} details our required assumptions and resulting inference. Section \ref{simulations} examines performance via a simulation study and Section \ref{example} applies our procedure to two distinct HIV clinical trial datasets. Section
\ref{discussion} concludes the paper with a discussion of the work, limitations, and future directions. Table \ref{tab:summary} summarizes our notation and the proposed measures and estimates which we cover in the following sections.
\vspace*{-0.6cm}
\section{Resilience Measures  \label{proposed}}
\vspace*{-0.4cm}
\subsection{Setting and Proposed Measures}

As described above, our setting is such that in Study A, we observe $\{S_{A0i}, Y_{A0i}\}$ for subjects $i = 1,...,n_{A0}$ from the control group and $\{S_{A1i}, Y_{A1i}\}$ for subjects $i = 1,...,n_{A1}$ from the treatment group. Let 
$\mu_{Ag}(s) \equiv E(Y\supg_{A}| S\supg_{A}=s)$, the true conditional mean functions in Study A. In Study B,  we only observe $\{S_{B0i}\}$ for subjects $i = 1,...,n_{B0}$ in the control group and $\{S_{B1i}\}$ for subjects $i = 1,...,n_{B1}$ in the treatment group. Let 
$\mu_{Bg}(s) \equiv E(Y\supg_{B}| S\supg_{B}=s)$, the true conditional mean functions in Study B. We are interested only in the setting, where 
$S\supone_{B}$ is stochastically greater than $S\supzero_{B}$. The ultimate parameter of interest is the average  treatment effect on $Y$ in Study B, denoted as  $$\Delta_B = E(Y_B\supone - Y_B\supzero),$$ which is not directly estimable since $Y$ is not measured in Study B. The surrogate paradox occurs when $\Delta_B<0$. Our goal is to understand to what extent and in what circumstances we can guarantee that $\Delta_B\geq 0$ i.e., that the paradox does not occur in Study B. 

Note that $\Delta_B$ can be expressed as  
$$ \Delta_B = \Delta_B(\mu_{B1}, \mu_{B0})  \equiv E\{\mu_{B1}(S^{(1)}_B)-\mu_{B0}(S^{(0)}_B)\},$$
where the expectation is with respect to $(S_B^{(1)}, S_B^{(0)}),$ and thus the surrogate paradox occurs, when $\Delta_B(\mu_{B1}, \mu_{B0})<0.$ Since we do not know $(\mu_{B1}, \mu_{B0})$, with a slight abuse of notation, for any pair of functions $(\mu_1^*, \mu_0^*)$ representing possible conditional mean functions, we define $$ \Delta_B(\mu_1^*, \mu_0^*) \equiv E\{\mu_1^*(S^{(1)}_B)-\mu_0^*(S^{(0)}_B)\}.$$ 
If $(\mu_1^*, \mu_0^*) = (\mu_{B1}, \mu_{B0}),$ then $ \Delta_B(\mu_1^*, \mu_0^*) = \Delta_B$, but we do not assume this to be true. Suppose for a moment that $S_B^{(1)}$ and $S_B^{(0)}$ follow known distributions with the CDFs being $F_{S_B^{(1)}}(\cdot)$ and $F_{S_B^{(0)}}(\cdot),$ respectively. In such a case, 
$$\Delta_B(\mu_1^*, \mu_0^*)=\int \mu_1^*(s)dF_{S_B^{(1)}}(s)-\int \mu_0^*(s)dF_{S_B^{(0)}}(s).$$
 Consequently, when treating $(\mu_1^*, \mu_0^*)$ as a pair of random functions, $\Delta_B(\mu_1^*, \mu_0^*)$ becomes a random variable. We propose to consider the probability that the surrogate paradox occurs with respect to a particular distribution of $(\mu_1^*, \mu_0^*)$ as: 
$$p_0=P\left(\Delta_B(\mu_1^*, \mu_0^*)<0\right),$$
which we refer to as the \textit{resilience probability}. Importantly, this probability is with respect to $(\mu_1^*, \mu_0^*).$ In addition, we consider the $\alpha$th quantile of $\Delta_B(\mu_1^*, \mu_0^*),$ $q_\alpha,$ such that 
$$ P(\Delta_B(\mu_1^*, \mu_0^*)>q_\alpha)=1-\alpha.$$ We refer to $q_\alpha$ as the \textit{resilience bound}. For some small given $\alpha,$ one may state that there is a high confidence that the average treatment effect on $Y$ is greater than $q_\alpha.$ 

Given the distribution of $(\mu_1^*, \mu_0^*),$ both $q_\alpha$ and $p_0$ can be obtained analytically or by simulating a large number of $(\mu_1^*, \mu_0^*)$ and $\Delta_B(\mu_1^*, \mu_0^*).$  Clearly, $q_\alpha$, $p_0$ and the distribution of $\Delta_B(\mu_1^*, \mu_2^*)$ all depend on the distribution of $\mu_1^*$ and $\mu_0^*.$  Of course, we  know neither the true distributions of $S_B^{(1)}$ and $S_B^{(0)}$ (we only have observed values) nor the distribution of $(\mu_1^*, \mu_0^*)$ (we only have the estimated conditional mean functions from Study A). Given this, in the following sections, we discuss how one may estimate these quantities with the observed data.

\subsection{Estimation of the Resilience Probability and Bound \label{interval}}

Let us first address the issue of $S_B^{(1)}$ and $S_B^{(0)}$ and temporarily assume that the distribution of $(\mu_1^*, \mu_0^*)$ is known, which we will relax. Instead of the true distributions of $S_B^{(1)}$ and $S_B^{(0)}$, we have observed values $\{S_{B0i}, i=1, \cdots, n_{B0}\}$ in the control group and $\{S_{B1i}, i = 1,...,n_{B1}\}$ in the treatment group. For a given $(\mu_1^*, \mu_0^*)$, we may first approximate $\Delta_B$ by 
$$\widehat{\Delta}_B(\mu_1^*, \mu_0^*)=\frac{1}{n_{B1}}\sum_{i=1}^{n_{B1}} \mu_1^*(S_{B1i})-\frac{1}{n_{B0}}\sum_{i=1}^{n_{B0}} \mu_0^*(S_{B0i}).$$

\noindent and for a given distribution of $(\mu_1^*, \mu_0^*),$ we may estimate $p_0$  and $q_\alpha$ by 

$$\widehat{p}=P\left\{\widehat{\Delta}_B(\mu_1^*, \mu_0^*) < 0\right\}$$
and $\widehat{q}_\alpha,$ the $\alpha$th percentile of $\widehat{\Delta}_B(\mu_1^*, \mu_0^*),$ respectively. 

Now we address the fact that we do not know $(\mu_1^*, \mu_0^*)$ and the distribution thereof. Although we have Study A, which we can use to estimate $(\mu_{A1}, \mu_{A0})$, denoted as $\widehat{\mu}_1(s)$ and $\widehat{\mu}_0(s)$, we deliberately do not want to assume that Study B functions are identical to those from Study A. Instead, we are willing to assume that $(\mu_1^*, \mu_0^*)$ is a member of a class of functions, whose deviations from $\widehat{\mu}_0(s)$ and $\widehat{\mu}_1(s)$ are stochastically controlled by a finite set of parameters $\bPi$. (We formalize this assumption in Assumption \ref{asmp:2} of Section \ref{inference}.) Our central idea is to repeatedly sample pairs $(\mu_1^*, \mu_0^*)$ from this class, compute the corresponding $\widehat{\Delta}_{Bj}=\widehat{\Delta}_B(\mu_1^*, \mu_0^*)$ for $ j=1, \cdots, J,$ use the empirical distribution of $\widehat{\Delta}_{Bj}, j=1,\cdots, J$ to approximate the distribution of $\Delta_B(\mu_0^*, \mu_1^*)$, and then use it to estimate the resilience probability, $p_0$, and resilience bound, $q_{\alpha}$. Figure \ref{fig:GP_fixed_HIV}, which we describe in more detail in our data application, illustrates the idea of generating these $\mu_1^*, \mu_0^*$ functions, shown as gray lines, as deviations from the estimated  $\widehat{\mu}_0(s)$ and $\widehat{\mu}_1(s)$, shown in black; he pink lines highlight one iteration where the generated conditional mean functions, $(\mu_1^*, \mu_0^*)$, result in a negative treatment effect estimate.

In Algorithm \ref{general_alg}, we describe this general algorithm without yet including details on how to exactly generate $\mu_0^*$ and $\mu_1^*$. In Step 1, we estimate the conditional mean functions $(\widehat{\mu}_1, \widehat{\mu}_0)$ using Study A. In Steps 2–5, we generate candidate pairs $(\mu_1^*, \mu_0^*)$ as controlled deviations from $\widehat{\mu}_1(s)$ and $\widehat{\mu}_0(s)$ and compute the corresponding $\widehat{\Delta}_B(\mu_0^*, \mu_1^*)$ using the observed surrogates in Study B. Repeating this process $J$ times (e.g., $J=500$) yields an empirical distribution $ \left\{\widehat{\Delta}_{B j} \mid j=1, \cdots J \right\}.$ Finally, in Steps 7-8, we use this empirical distribution to obtain $\widehat{p}$ and  $\widehat{q}_{\alpha}$ as summary functionals. Now that we have described the general algorithm, it will be helpful to explicitly express our proposed quantities in terms of the cumulative distribution function (CDF) of $\Delta_B(\mu_1^*, \mu_0^*)$, where $(\mu_1^*, \mu_0^*)$ follows a distribution governed by $\bPi$. We denote the CDF of $\Delta_B(\mu_1^*, \mu_0^*)$ as $F_{\bPi}$ and the empirical CDF, obtained using $ \left\{\widehat{\Delta}_{B j} \mid j=1, \cdots J \right\}$ as $\widehat{F}_{\bPi}$. The true and estimated resilience probability can then be expressed as $p_0 = F_{\bPi}(0)$ and $\widehat{p} = \widehat{F}_{\bPi}(0)$, respectively.  The true and estimated resilience bound can be expressed as $q_{\alpha} = F_{\bPi}^{-1}(\alpha)$ and $\widehat{q}_{\alpha} = \widehat{F}_{\bPi}^{-1}(\alpha)$, respectively.

\begin{algorithm}[ht] 
\caption{General Algorithm } \label{general_alg}
\begin{algorithmic}[1]
    \State Estimate $\mu_0(s)$ and $\mu_1(s)$ in Study A, denoted by  $\widehat{\mu}_0$ and $\widehat{\mu}_1$.  For example, this can be achieved by using the nonparametric Nadaraya-Watson kernel estimator \citep{nadaraya1964estimating,watson1964smooth}:
\begin{equation*}
\widehat{\mu}_g(s) =\frac{\sum_{i=1}^{n_{Ag}}K_{h_g}(S_{Agi}-s)Y_{Agi}}{\sum_{i=1}^{n_{Ag}}K_{h_g}(S_{Agi}-s)}
\end{equation*}
where $K(\cdot)$ is a smooth symmetric density function with finite support, $K_h(\cdot)=K(\cdot/h)/h$, and $h_g$ is the bandwidth where $h_g=O_p(n_g^{-\delta})$, $1/5<\delta<1/3$, which is a consistent estimator of $\mu_{Ag}(s)$ \citep{mack1982weak}. 
    \For{$j = 1,...,J$}
        \State Generate new $\mu_0^*$ and $\mu_1^*$ within a general class of functions characterized as deviations from $\widehat{\mu}_0(s)$ and $\widehat{\mu}_1(s)$ governed by parameters $\bPi$. 
         \State Use the new $\mu_0^*$ and $\mu_1^*$ to generate realizations $\widehat{y}_{B1i} = \mu_0^*(S_{B1i})$ for $i = 1,...,n_{B0}$ and $\widehat{y}_{B0i} = \mu_1^*(S_{B0i})$ for $i = 1,...,n_{B1}$.
        \State Calculate $\widehat{\Delta}_{Bj} = n_{B1}^{-1}\sum_{i=1}^{n_{B1}} \widehat{y}_{B1i} - n_{B0}^{-1}\sum_{i=1}^{n_{B0}} \widehat{y}_{B0i}$.
    \EndFor
 \State Estimate the resilience probability, $p_0$, as:
\begin{equation*}
    \widehat{p} = J^{-1} \sum_{j-1}^J I(\widehat{\Delta}_{B j} <0).
\end{equation*}
    \State Estimate the resilience bound, $q_{\alpha}$, as the $\alpha$th percentile of $ \{\widehat{\Delta}_{B j} \}_{j=1}^{J}$.
\begin{equation*}
    \widehat{q}_{\alpha} = \inf \left\{ q \in \mathbb{R} : \frac{1}{J} \sum_{j=1}^{J} I(\widehat{\Delta}_{B j} \leq q) \geq \alpha \right\}.
\end{equation*}
\end{algorithmic}
\end{algorithm}

Next, we define more refined versions of this general algorithm with a specific class of functions selected for generating $\mu_0^*$ and $\mu_1^*$: one based on Gaussian processes, one based on polynomial functions, and one based on Fourier series. The unifying theme for all three versions is that the assumed class of functions uses the estimated conditional mean functions from Study A as their ``center". For each version, there are parameters that govern the characterization of the class, $\bPi$, which control the deviations from the Study A estimates; we fix these parameters throughout, with further discussion in Section \ref{discussion}. 

\subsubsection{Resilience Probability and Bound within a Gaussian Process Class} 
Our first refinement models the class of functions as Gaussian processes (GPs) and is described in Algorithm \ref{unified_alg}. Specifically, we assume that $(\mu_1^*, \mu_0^*)$ come from Gaussian processes:  
$$\mu_0^*(s) \sim GP(\eta_0(s), k_0(s,s')) \mbox{ and }\mu_1^*(s) \sim GP(\eta_1(s), k_1(s,s'))$$
where $\eta_0(s)$ and $\eta_1(s)$ are smooth mean functions, and $k_0(s,s')$ and $k_1(s,s')$ are covariance kernels. The smooth mean functions $\eta_0(s)$ and $\eta_1(s)$ are set to be the nonparametric conditional mean estimates from Study A, $\widehat{\mu}_0(s)$ and $\widehat{\mu}_1(s)$. We assume an RBF (radial basis function) kernel of the form:
$$k_g(s,s') = \sigma_g^2\exp\left(-\frac{(s-s')^2}{2\theta_g^2}\right).$$
That is, we assume $\mu_0^*(s)$ and $\mu_1^*(s)$ are structured deviations captured using a Gaussian Process specification and the parameters governing these deviations are contained in $\bPi=\{\sigma_1, \sigma_0, \theta_1, \theta_0\}.$  We assume $\sigma \equiv \sigma_1=\sigma_0$ and $\theta \equiv \theta_1=\theta_0$, though this assumption can be relaxed if needed, and fix these parameters at specified values.  

\begin{algorithm}[h]
\caption{Class-specific Algorithm: Gaussian Process, Polynomial, Fourier Series} \label{unified_alg}
\begin{algorithmic}[1]
    \State Estimate $\mu_{A0}(s)$ and $\mu_{A1}(s)$ in Study A, denoted by  $\widehat{\mu}_0(s)$ and $\widehat{\mu}_1(s)$, using the nonparametric Nadaraya-Watson kernel estimator.
    \State Select the function class $\mathcal{F}$: one of Gaussian Process (GP), Polynomial (Poly), or Fourier Series (Fourier).
    \State Select or estimate the parameters, $\bPi$, for $\mathcal{F}$, such as the covariance kernel parameters or covariance matrix $\bSigma$. Let $\bS_{Bg}$ indicate the vector of $S_{Bgi}$ for $i = 1,...,n_{Bg}$ for $g=0,1$.
    \For{$j = 1,...,J$}
        \If{$\mathcal{F} = $ GP}
            \State Define $\widehat{\mu}_g(\bS_{Bg})$ and covariance matrices $\widehat{k}_g(\bS_{Bg}, \bS_{Bg})$.
            \State Generate $\widehat{\by}_{Bg} \sim MVN(\widehat{\mu}_g(\bS_{Bg}), \widehat{k}_g(\bS_{Bg}, \bS_{Bg}))$.
        \ElsIf{$\mathcal{F} = $ Poly}
            \State Generate coefficients $\bbeta_g \sim N(0, \bSigma)$ for $g=0,1$.
            \State Compute $\widehat{\by}_{Bg} = \widehat{\mu}_{g}(\bS_{Bg}) + \sum_{j=1}^{d-1} \beta_{g,j} \left(z_{Ag}^{-1}(\bS_{Bg} - \bar{s}_{Ag})\right)^d$.
        \ElsIf{$\mathcal{F} = $ Fourier}
            \State Generate $\bbeta_g \sim N(0, \bSigma)$ for $g=0,1$.
            \State Compute $\widehat{\by}_{Bg} = \widehat{\mu}_g(\bS_{Bg}) + \beta_{0,0} + \sum_{j=1}^{d-1} \beta_{0, j} \left[\sin\left(\frac{\bS_{Bg} - c_g}{B_j}\right) + \cos\left(\frac{\bS_{Bg} - c_g}{B_j}\right)\right]$.
        \EndIf
        \State Compute $\widehat{\Delta}_{Bj} = n_{B1}^{-1}\sum_{i=1}^{n_{B1}} \widehat{y}_{B1i} - n_{B0}^{-1}\sum_{i=1}^{n_{B0}} \widehat{y}_{B0i}$.
    \EndFor
    \State Estimate the resilience probability:
    \begin{equation*}
        \widehat{p} = J^{-1} \sum_{j=1}^J I(\widehat{\Delta}_{B j} < 0).
    \end{equation*}
    \State Estimate the resilience bound:
    \begin{equation*}
        \widehat{q}_{\alpha} = \inf \left\{ q \in \mathbb{R} : \frac{1}{J} \sum_{j=1}^{J} I(\widehat{\Delta}_{B j} \leq q) \geq \alpha \right\}.
    \end{equation*}
\end{algorithmic}
\end{algorithm}

\subsubsection{Resilience Probability and Bound within a Polynomial Function Class} 

Our second refinement models the class of functions as general polynomials and is described in Algorithm \ref{unified_alg}. We considered polynomials because of their ability to flexibly approximate arbitrary functions \citep{rudinprinciples}. Specifically,  we assume that 
$$\mu_0^*(s) =  \widehat{\mu}_0(s) + \sum_{j=0}^{d-1} \beta_{0,j} \cdot \left(\frac{s - \bar{s}_{A0}}{z_{A0}} \right)^j$$
$$\mu_1^*(s) = \widehat{\mu}_1(s) + \sum_{j=0}^{d-1} \beta_{1,j} \cdot \left( \frac{s - \bar{s}_{A1}}{z_{A1}} \right)^j$$ 
where $\bar{s}_{Ag}$ and $z_{Ag}$ are the sample mean and standard deviation, respectively, of the observed $S_{Ag}$ used to standardize $s$; and 
$\bbeta_g \sim MVN(0, \bSigma)$, $\bbeta_g = (\beta_{g,0}, \beta_{g,1},...,\beta_{0,d-1})'$ and $\bSigma$ is a diagonal matrix. That is, we assume $\mu_0^*(s)$ and $\mu_1^*(s)$ are structured deviations captured using polynomial expansions of degree $d-1$ (we use $d=4$) and the parameters governing these deviations are contained in $\bSigma$ i.e., $\bPi=\{\bSigma\}$ within this algorithm. Similar to the prior algorithm, we fix these parameters at specified values.

\subsubsection{Resilience Probability and Bound within a Fourier Series Class} 

Our last refinement models the class of functions as Fourier series and is described in Algorithm \ref{unified_alg}. We considered Fourier series given the fact that any sufficiently well-behaved continuous function on a bounded interval can be approximated by a sum of sine and cosine functions with varying frequencies \citep{rudinprinciples}. Here, we assume that 
$$\mu_0^*(s) =  \eta_0(s) + \beta_{0,0} + \sum_{j=1}^{d-1} \beta_{0, j} \left[\sin\left(\frac{s - c_0}{B_j}\right) + \cos\left(\frac{s - c_0}{B_j}\right)\right]$$
$$\mu_1^*(s) =  \eta_1(s) + \beta_{1,0} + \sum_{j=1}^{d-1} \beta_{1, j} \left[\sin\left(\frac{s - c_1}{B_j}\right) + \cos\left(\frac{s - c_1}{B_j}\right)\right],$$
where $\bbeta_g \sim MVN(0, \bSigma)$, $\bbeta_g = (\beta_{g,0}, \beta_{g,1},...,\beta_{0,d-1})'$ and $\bSigma$ contains the parameters governing the deviations, $\bSigma$ i.e., $\bPi=\{\bSigma\}$,  and  $B_1,...,B_{d-1}$ are given constants that control the period of the resulting function. 
In our implementation, we use $d=4$ and fix $c_0 = \min(S_{A0i})$ and $c_1 = \min(S_{A1i})$. In addition, similar to the prior algorithms, we fix the parameters of $\bSigma$ at specified values. 

In summary, we have proposed a general algorithm as well as three specific algorithms to estimate the resilience probability and resilience bound. In Section \ref{inference}, we detail our required assumptions and resulting inference, and in Section \ref{simulations} we examine the finite sample properties of the estimates. However, before doing so, in the following section we propose and describe the resilience set, our third and final resilience measure.  
\vspace*{-0.6cm}
\subsection{The Resilience Set \label{res_set}}

The resilience probability $p_0$ and the resilience bound $q_{\alpha}$ are defined with respect to a specific distribution of $(\mu_1^*, \mu_0^*)$. In contrast, it may also be of interest to explicitly identify \textit{all possible distributions} of $(\mu_1^*, \mu_0^*),$ such that the corresponding $q_\alpha$ is sufficiently large, that is,

$$\Omega_q = \{F \mid q_\alpha(F) \geq 0\}.$$
In other words, $\Omega_q$ represents the set of all possible distributions $F$ of $(\mu_1^*, \mu_0^*)$ such that the corresponding resilience bound $q_{\alpha}$ is non-negative. We refer to $\Omega_q$ as the \textit{resilience set}. Given our restriction to distributions of $(\mu_1^*, \mu_0^*)$ that belong to a class of functions with deviations governed by $\bPi$, we re-define the resilience set to make this restriction explicit and, with a slight abuse of notation, denote the resilience bound $q_{\alpha} = F_{\bPi}^{-1}(\alpha)$ as $q_\alpha(F_{\bPi})$. Specifically, 
$$\Omega_q \equiv \{\bPi \mid q_\alpha(F_{\bPi}) \geq 0\},$$
the set of all parameters $\bPi$ such that the corresponding resilience bound $q_\alpha(F_{\bPi})$ is non-negative.  Naturally, the form of $\Omega_q$ depends on the assumed class: for the Gaussian Processes class, $\bPi = (\sigma^2, \theta)$, and for the polynomial or Fourier series class, $\bPi = \{\bSigma_{d \times d}\}$, which may correspond to arbitrary positive semi-definite matrices, diagonal matrices, or scaled identity matrices of the form $\sigma^2 I_{d \times d}$.

In Appendix A, we derive closed-form expressions for $F_{\bPi}$ for each function class considered. For example, within the Gaussian Process class, $ P(\Delta_B(\mu_1^*, \mu_0^*) < 0)$ is approximately normally distributed, with mean and variance given as explicit functions of $\sigma^2$ and $\theta$. Thus, the quantity $P(\Delta_B(\mu_1^*, \mu_0^*) < 0)$ can be expressed directly in terms of $\sigma^2$ and $\theta$, and this normal approximation can be used to identify regions of the parameter space, where $q_\alpha \geq 0$. We define the estimated resilience set as:
$$\widehat{\Omega}_q =  \{\bPi \mid \widehat{q}_\alpha(\widehat{F}_{\bPi}) \geq 0\},$$  
where $\widehat{q}_\alpha = \widehat{F}^{-1}_{\bPi} (\alpha)$ is denoted as $\widehat{q}_\alpha(\widehat{F}_{\bPi}).$ To identify this set in practice, we proceed in two steps. First, we construct a dense grid of candidate values for $\bPi$, compute the corresponding resilience bound $\widehat{q}_\alpha(\widehat{F}_{\bPi})$ for each grid point, and retain the set of parameters satisfying $\widehat{q}_\alpha(\widehat{F}_{\bPi}) \geq 0$. Second, we use numerical optimization to more precisely characterize the boundary of the resilience set. Figure \ref{fig:aids_resilience_sets}, described in more detail in Section \ref{example}, displays the resilience set for the HIV data application using the Gaussian Process algorithm parameterized by $\sigma^2$ and $\theta$, with an $\alpha=0.10$, where blue shading indicates the identified pairs, $(\theta,\sigma^2)$, such that $\widehat{q}_\alpha(\widehat{F}_{\bPi}) \geq 0.10$ when using the grid search and the solid black line indicates the boundary of these pairings using numeric optimization. Informally, this means that, given a Study A and a Study B as described in our setting, a desired level $\alpha$, and a selected function class, we can return an estimated subset of the parameter space of $\bPi$ such that the estimated probability of the surrogate paradox is small. We illustrate this proposed approach using each algorithm in our data application in Section \ref{example}.

\vspace*{-0.6cm}
\section{Assumptions and Inference  \label{inference}}
There are two important approximations inherent in this procedure. First, we condition on the observed surrogates in Study B, $S_{B0i}$ and $S_{B1i}$, treating them as random realizations from respective distributions. Second, we use estimates $(\widehat{\mu}_1(s), \widehat{\mu}_0(s))$ based on Study A to characterize the possible functions for $\mu_{Bg}(s), g=0, 1,$ in Study B, and treat them as fixed. In this section, we formalize these points as well as our imposed assumptions, which include our restriction of $(\mu_1^*, \mu_0^*)$ to a class of functions with deviations governed by parameters $\bPi$, and detail the resulting appropriate inference. Specifically, we impose the following assumptions:

\begin{assumption}[Consistency]\label{asmp:c}
For each subject $i$ in Study $K \in \{A, B\}$, if the subject receives treatment $g \in \{0,1\}$, then the observed outcomes equal the corresponding potential outcomes: $S_{Kgi} = S_K^{(g)}$ and $Y_{Agi} = Y_A^{(g)}$.
\end{assumption}

\begin{assumption}[Positivity]\label{asmp:p}
The probability of receiving each treatment is strictly positive: $0 < P(G_K=1) < 1$, where $G_K$ is the treatment assignment indicator in Study $K$.
\end{assumption}

\begin{assumption}[Unconfoundedness]\label{asmp:u}
Treatment assignment is independent of potential outcomes: $(Y_K^{(0)}, S_K^{(0)}, Y_K^{(1)}, S_K^{(1)}) \perp G_K$ for $K \in \{A, B\}$.
\end{assumption}

\begin{assumption}[Independent and Identically Distributed Observations]\label{asmp:1}
Within Study B, the observed surrogate outcomes in the control group, $\{S_{B0i}\}_{i=1}^{n_{B0}}$, are independent and identically distributed (i.i.d.), and the observed surrogate outcomes in the treatment group, $\{S_{B1i}\}_{i=1}^{n_{B1}}$, are i.i.d. Furthermore, the sample proportions satisfy $n_{B0}/(n_{B0}+n_{B1}) \rightarrow r \in (0,1)$ as $n_{B0}, n_{B1} \to \infty$.
\end{assumption}


\begin{assumption}[Basis Expansion of Conditional Mean Functions]\label{asmp:2}
The true conditional mean functions in Study B can be represented as a finite basis expansion:
$$\mu_{Bg}(s) = \sum_{k=1}^K W_{gk} C_k(s),$$
for $g \in \{0,1\}$, where $\{C_k(\cdot)\}_{k=1}^K$ are known bounded basis functions, and $\{W_{gk}\}$ are random coefficients with a joint distribution governed by a finite set of parameters $\bPi \in \mathcal{P}$, with continuous density and finite second moments.
\end{assumption}

\begin{assumption}[Regularity Conditions]\label{asmp:regularity}
We assume (i) $F_{\bPi}$ is continuous and strictly increasing at $q_\alpha$, and the corresponding density $f_{\bPi}$ is continuous at $q_\alpha$ and $\inf_{\bPi\in {\cal P}}f_{\bPi}(q_\alpha)>0$; 
(ii) $F_{\bPi}(q)$ can be written in the form of $h(\theta, \bPi, q),$ where $h(\theta, \bPi, q)$ is a differentiable function and $\theta$ is a vector given in Appendix B, and the class of functions $\{h(\cdot, \bPi, q)\mid \bPi\in {\cal P}, q\}$ is Glivenko-Cantelli; (iii) the parameter space $\mathcal{P}$ for $\bPi$ is compact. 
\end{assumption}

Assumptions \ref{asmp:c}-\ref{asmp:u} are standard conditions for causal inference under the potential outcomes framework, ensuring that treatment effects are well-defined and identifiable. Assumption \ref{asmp:1} imposes standard regularity conditions on the surrogate measurements in Study B to guarantee convergence of empirical quantities. 
Assumption \ref{asmp:2} requires that the unknown conditional mean functions in Study B, $\mu_{B0}(s)$ and $\mu_{B1}(s)$, can be expressed in a basis function representation with random coefficients. This structure provides a flexible yet tractable framework for modeling potential deviations from the conditional means estimated in Study A. We verify in Appendix A that the structure imposed in each of the proposed algorithms (Gaussian Process, polynomial, and Fourier) in Section \ref{interval} satisfy this assumption. Lastly, we consider that $(\widehat{\mu}_{1}(s), \widehat{\mu}_{0}(s))$ is sufficiently close to $(\mu_{A1}(s), \mu_{A0}(s))$ to serve as the ``center'' of the distribution of $(\mu_1^*, \mu_0^*)$ for characterizing the transportability from $(\mu_{A1}(s), \mu_{A0}(s))$ to $(\mu_{B1}(s), \mu_{B0}(s)).$ We discuss these assumptions further in Section \ref{discussion}.

Next, we investigate the asymptotic properties of each proposed estimate; the following asymptotic results are stated in terms of $n_{B1}, n_{B0} \to \infty$. 
\begin{theorem}[Asymptotic Properties of $\widehat{p}$]\label{thm:asymp_normality}
Under Assumptions \ref{asmp:c}--\ref{asmp:regularity}, the estimator $\widehat{p}$ is consistent for $p_0$, and satisfies
$$
\sqrt{\frac{n_{B1}n_{B0}}{n_{B1}+n_{B0}}}(\widehat{p} - p_0) \xrightarrow{d} N(0, \sigma_P^2),$$
where $\sigma_P^2$ is the asymptotic variance defined in Appendix B.
\end{theorem}

\begin{theorem}[Asymptotic Properties of $\widehat{q}_\alpha$]\label{thm:qalpha_normality}
Under Assumptions \ref{asmp:c}--\ref{asmp:regularity}, the estimator $\widehat{q}_\alpha$ is a consistent estimator of $q_\alpha$ and
$$\sqrt{\frac{n_{B1}n_{B0}}{n_{B1}+n_{B0}}}(\widehat{q}_\alpha - q_\alpha) \xrightarrow{d} N(0, \sigma_Q^2), $$
where $\sigma_Q^2$  is the asymptotic variance defined in Appendix B.
\end{theorem}
\noindent The asymptotic properties of $\widehat{\Omega}_q$ differ slightly from those of standard estimators because $\widehat{\Omega}_q$ is a set. Therefore, we describe the consistency of $\widehat{\Omega}_q$ in terms of convergence to the true set $\Omega_q$ as the sample size increases, using the outer Hausdorff distance as the notion of convergence, formalized in Theorem \ref{thm:asymp_set}. 

\begin{theorem}[Consistency of $\widehat{\Omega}_q$] \label{thm:asymp_set}
Under Assumptions \ref{asmp:c}--\ref{asmp:regularity}, the estimated resilience set $\widehat{\Omega}_q$ is consistent for $\Omega_q$ in the sense that
$$d_H(\widehat{\Omega}_q, \Omega_q) \xrightarrow{p} 0,$$
\noindent where $d_H(T,U)$ denotes the Hausdorff distance between two sets $T$ and $U$:
$$d_H(T,U) = \inf\left\{ \delta > 0 : T \subseteq U^\delta \text{ and } U\subseteq T^\delta \right\},$$
where $U^\delta = \{ x \in \mathbb{R}^d : \inf_{u \in U} \|x - u\| \leq \delta \}$ is the $\delta$-expansion of $U$, and similarly for $T$.
\end{theorem}

\noindent Proofs of Theorems \ref{thm:asymp_normality} - \ref{thm:asymp_set} are provided in Appendix B. For variance estimation in our numerical studies that follow, we use the nonparametric bootstrap to estimate $\sigma_P$ and $\sigma_Q$.

 \vspace*{-0.6cm}
\section{Simulation Study \label{simulations}}

\begin{landscape}
\begin{figure}[ht]
    \centering
    \includegraphics[width=1.4\textwidth]{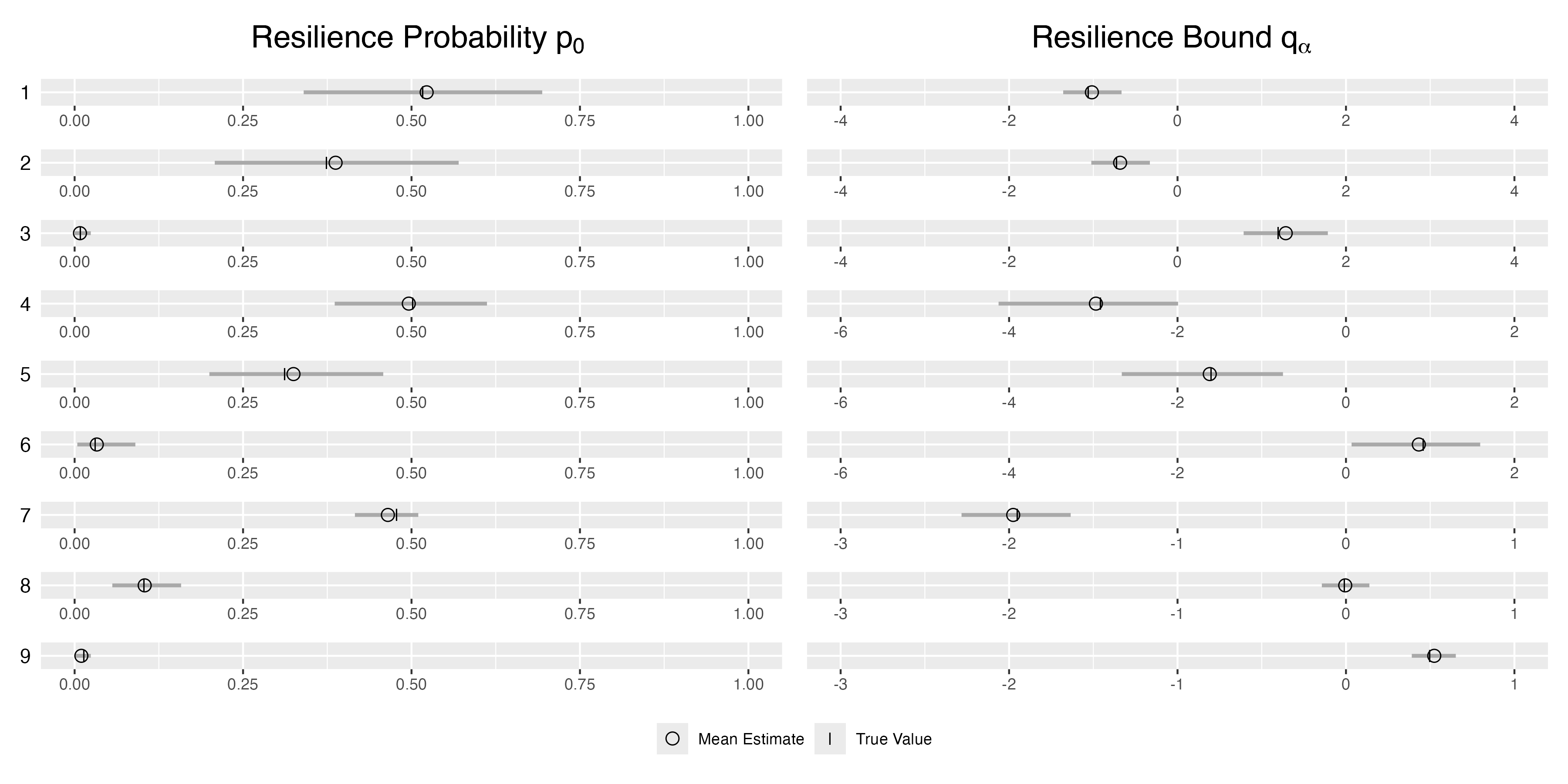}
    \caption{Simulation results estimating the resilience probability, $p_0$, and resilience bound $q_{\alpha}$ for $\alpha=0.10$ in Settings 1-9 via the Gaussian Process algorithm (in Settings 1-3), the polynomial algorithm (Settings 4-6), and the Fourier series algorithm (Settings 7-9); the open circle denotes the mean estimate, the vertical line denotes the truth, the grey horizontal line represents the interval between the 2.5th and 97.5th percentiles of the estimates, capturing the central 95\% of the distribution. Note that for each setting, the algorithm used for estimation corresponded to the data generation (e.g. Gaussian Process algorithm was used for estimation in Settings 1-3); in Appendix C we additionally investigate robustness to misspecification of the basis class in these settings.}
    \label{fig:all_plots_together}
\end{figure}
\end{landscape}

We evaluated the finite-sample performance of our proposed methods to estimate the resilience probability and resilience bound using a simulation study. We investigated nine simulation settings; full details on and the motivation for these settings are provided in Appendix C. Briefly, in Settings 1-3, the data were generated such that the Gaussian process function specification was correct. In Settings 4-6, the data were generated such that the polynomial function specification was correct. In Settings 7-9, the data were generated such that the Fourier series function specification was correct. In Settings 1, 4, and 7, the occurrence of the surrogate paradox in Study B was probable, defined as $p_0 \in (0.47,0.52$); in Settings 2, 5, and 8, the occurrence of the surrogate paradox in Study B was possible, defined as $p_0 \in (0.10,0.38)$; in Settings 3, 6, and 9, the occurrence of the surrogate paradox in Study B was unlikely, defined as $p_0 \in (0,0.04)$. Simulation results summarize 1000 replications with $n_A=800$ ($400$ in each group) and $n_0 =400$ ($200$ in each group, similar to our sample sizes in the application). Table \ref{tab:gp_fixed_p} and Figure \ref{fig:all_plots_together} display the simulation results. Across all settings, the algorithms demonstrate reasonable performance with small bias for estimating $p_0$ and $q_{\alpha}$ and average standard estimates obtained via bootstrapping close to the empirical standard error estimates. Note that for each setting, the algorithm used for estimation corresponded to the data generation (e.g. Gaussian Process algorithm was used for estimation in Settings 1-3); in Appendix C we additionally investigate robustness to misspecification of the basis class.  
\vspace*{-0.6cm}
\section{Real Data Application \label{example}}
We illustrate our proposed methods using two HIV/AIDS clinical trial datasets examining various HIV treatments. For our illustration, Study A was the ACTG 320 study \citep{hammer1997controlled} and Study B was the ACTG 193A Study \citep{henry1998randomized}, both of which compared a two-drug arm (control group) to a three-drug arm (treated group). The primary outcome of interest was change in HIV-1 RNA per mL from baseline to 24 weeks and the surrogate marker was change in CD4 cell counts per cubic millimeter from baseline to 24 weeks, which is less costly to measure. Both RNA and CD4 cell counts (the primary outcome and the surrogate) were collected for all patients in Study A. In contrast, in Study B, CD4 cell count (the surrogate) was collected for all patients while RNA (the primary outcome) was not. Our analytic dataset in Study A had 412 individuals in the control group and 418 in the treated group; Study B had 176 in the control group and 176 in the treated group.

\begin{figure}[ht]
    \centering
    \includegraphics[width=\textwidth]{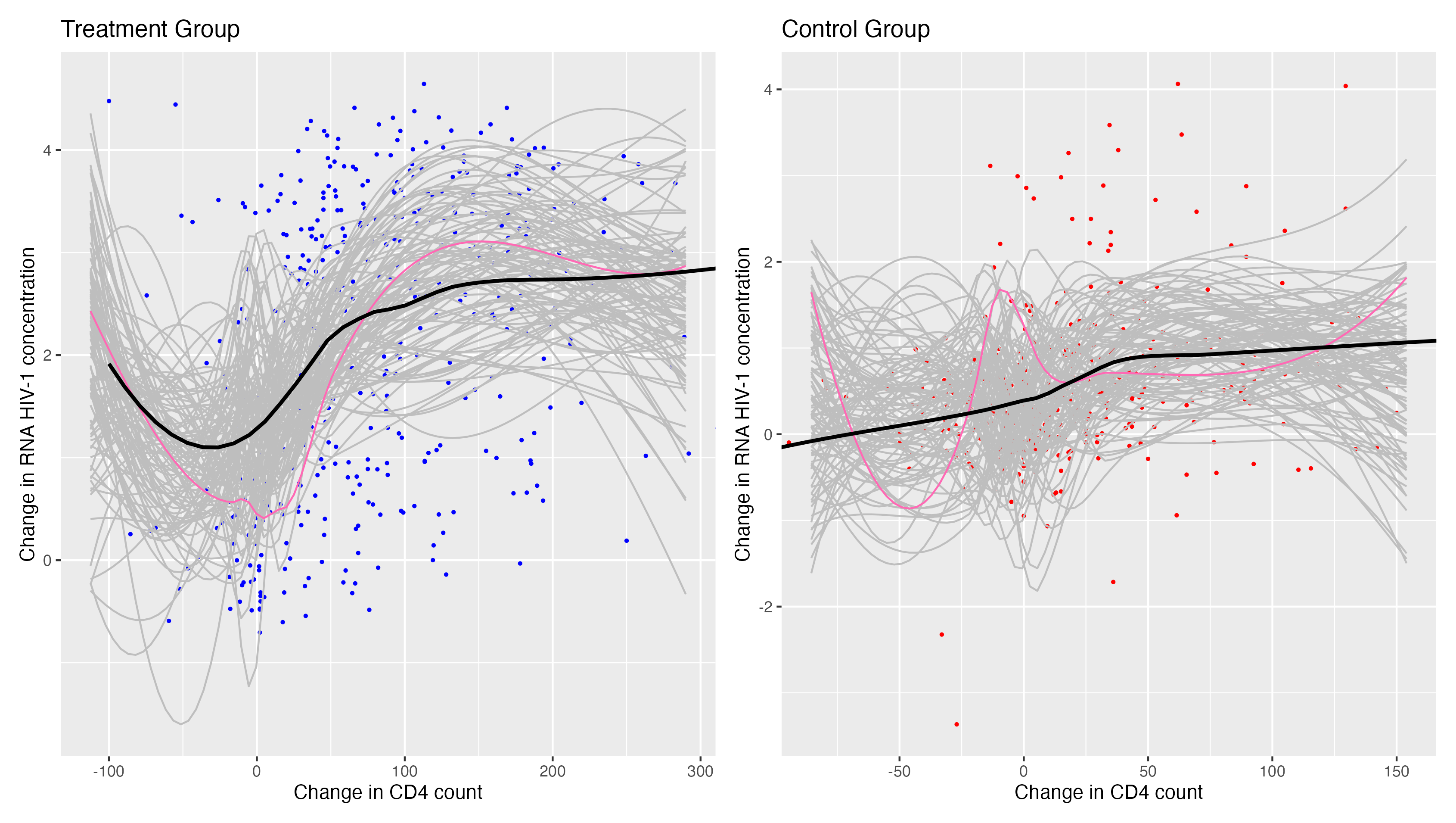}
    \caption{Data from the data application Study A, with treatment group data on the left and control group data on the right; the figure displays the observed data from Study A shown as individual points (blue for treatment, red for control), along with the kernel smoothed estimate of the conditional mean functions in each group (solid black line) and 100 iterations (out of 500) of the generated conditional mean functions for Study B (gray lines), generated and used by the Gaussian Process algorithm with  $\sigma^2 = 1$ and $\theta = 2$; the pink lines highlight one iteration where the generated conditional mean functions result in a negative treatment effect estimate; plots restricted to the support of the observed surrogates in Study B, used for estimation.}
    \label{fig:GP_fixed_HIV}
\end{figure}

\begin{figure}[ht]
    \centering
    \includegraphics[width=0.8\textwidth]{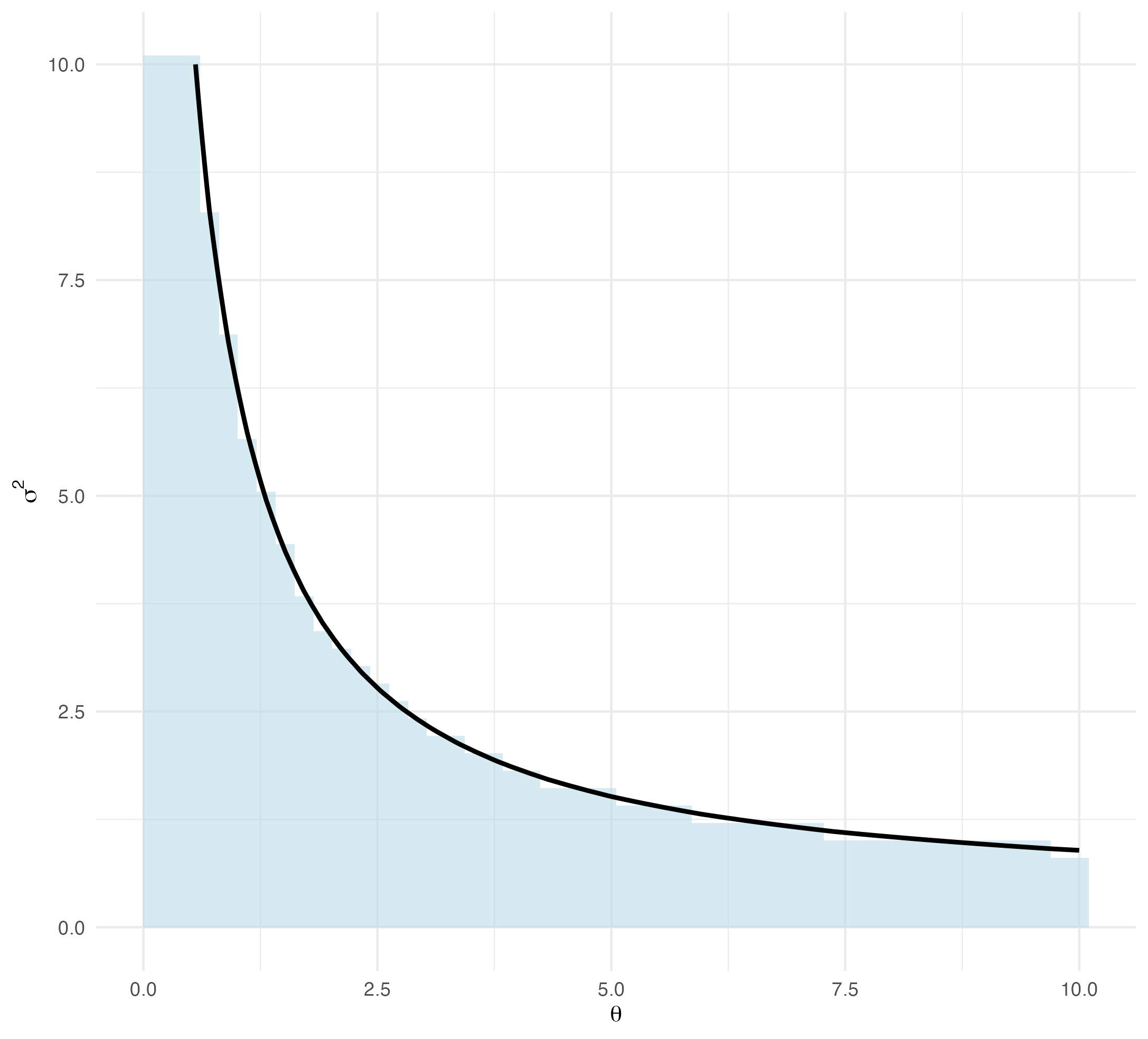}
    \caption{Estimated resilience set, $\widehat{\Omega}_q =  \{\bPi \mid \widehat{q}_\alpha(\widehat{F}_{\bPi}) \geq 0\}$, for the HIV data application using the Gaussian Process algorithm parameterized by $\sigma^2$ and $\theta$, with an $\alpha=0.10$; blue shading indicates the identified pairings of $\theta,\sigma^2$ where $\widehat{q}_\alpha(\widehat{F}_{\bPi}) \geq 0$ when using a grid search; the solid black line indicates the boundary of these pairings using numeric optimization.}
       \label{fig:aids_resilience_sets}
\end{figure}

We first applied our proposed approach using the Gaussian Process algorithm (with fixed $\sigma^2=1.25$ and $\theta=2$) to investigate resilience to the surrogate paradox. Figure~\ref{fig:GP_fixed_HIV} displays the observed data from Study A shown as individual points, along with the kernel smoothed estimate of the conditional means (solid black line) and 100 iterations (out of 500) of the generated conditional mean functions for Study B, used by the algorithm. Our approach resulted in an estimated resilience probability of $\widehat{p} = 0.014$, with a standard error (SE) of $0.0144$, and an estimated resilience bound of $q_{\alpha}=0.4077$, with $SE = 0.1136$, using $\alpha=0.10$. Because this estimated probability is relatively low, and the resilience bound is greater than 0, our results suggest that the risk of the surrogate paradox may be low. In addition, we estimate the resilience set, displayed in Figure \ref{fig:aids_resilience_sets}, where blue shading indicates the identified pairings of $\theta,\sigma^2$ where $\widehat{q}_\alpha(\widehat{F}_{\bPi}) \geq 0$ when using a grid search and the solid black line indicates the boundary of these pairings using numeric optimization. Practically, this means that if $\theta=1,\sigma^2=1.25$, for example, is a reasonable pairing to consider in terms of expected deviations of Study B from Study A, then the estimated resilience bound will be greater than 0 (because this pairing is contained in the blue shading), indicating a low risk of the surrogate paradox in Study B. In contrast, if $\theta=5.0,\sigma^2=5.0$, is a reasonable pairing to consider in terms of expected deviations of Study B from Study A, then the estimated resilience bound will be negative (because this pairing is \textit{not} contained in the blue shading), indicating a higher risk of the surrogate paradox in Study B.  In Appendix C, we provide additional results using the polynomial and Fourier Series algorithms.

\vspace*{-0.6cm}
\section{Discussion \label{discussion}}

In this work, we proposed resilience measures within a simulation-based framework to assess the robustness of using a surrogate marker to infer a treatment effect in a new study. Our approach defines and estimates a resilience probability, resilience bound, and resilience set, and provides tools for inference. Rather than assuming direct transportability of the conditional mean functions from one study to another, we introduced the idea of allowing the conditional mean function to deviate from the prior study in a controlled way, relying on a unified basis representation of the conditional mean functions. Though this unified basis representation encompasses several flexible modeling strategies, our imposition of this structure for the conditional mean functions is certainly a strong assumption. This is admittedly a limitation: the true conditional mean functions in Study B may not satisfy any particular basis representation. Fully nonparametric approaches to model these functions would be preferable in principle. However, in practice, full nonparametric modeling is infeasible here because Study B provides no observed outcomes $Y$, only surrogate values $S$. Thus, we cannot nonparametrically guess how $\mu_{B1}(s)$ and $\mu_{B0}(s)$ might differ in the new study without imposing some reasonable structure. Given this fundamental limitation, our approach represents, to our knowledge, the best available strategy to flexibly and plausibly model uncertainty about the conditional mean functions in Study B in this surrogate setting. Another important choice within our framework is the specification of the parameters within $\bPi$, for example, the $\sigma^2$ and $\theta$ in the Gaussian Process algorithm. In this work, we fixed these parameters at specified values, but alternative strategies could be considered. For instance, one could attempt to estimate both $\sigma^2$ and $\theta$ directly from the previous study data. Unfortunately, with only a single previous study, the joint estimation of these parameters is not clear. If multiple previous studies were available, hierarchical modeling strategies or empirical Bayes methods could be applied to estimate these hyperparameters across studies. One could certainly consider a fully Bayesian approach, placing priors on $\sigma^2$ and $\theta$ and sampling from the posterior distribution over outcome functions, a promising direction for future work.


Finally, an important question is how to use these resilience measures in practice. If the resilience probability is low or the resilience bound is unfavorable, we argue that this suggests that the surrogate marker is ``fragile", and that strong conclusions based on the surrogate in Study B should be reconsidered. In extreme cases, it may indicate that the surrogate marker should not be used at all to infer treatment effects in Study B. More generally, our framework provides a quantitative tool for assessing risk and uncertainty in surrogate-based analyses, and for informing more cautious interpretation when surrogate markers are applied in new populations or settings.

\vspace*{-0.6cm}
\section*{Data Availability} 
The data from the HIV clinical trials used in this paper are publicly available upon request from the AIDS Clinical Trial Group: \url{https://actgnetwork.org/submit-a-proposal}
\vspace*{-0.6cm}
\section*{Acknowledgments and Funding}
This work was supported by NIDDK grant R01DK118354 (PI:Parast). We are grateful to the AIDS Clinical Trial Group (ACTG) Network for supplying the ACTG data.
 
\vspace*{-0.6cm}
\section*{Supplementary material}
Online supplementary material includes all referenced appendices. The proposed methods are implemented in the R package \texttt{SurrogateParadoxTest} available at \url{https://github.com/emily13hsiao/SurrogateParadoxTest}, which also contains all code to reproduce all results.
\vspace*{-0.6cm}

\clearpage
\begin{table}[ht]
\caption{Summary of notation and defined resilience measures with corresponding descriptions \\} \label{tab:summary}
\centering
\renewcommand{\arraystretch}{1.3} 
\begin{tabular}{|l|>{\raggedright\arraybackslash}p{13cm}|}
\hline
\multicolumn{2}{|l|}{\textbf{Notation}} \\
\hline
Quantity & Description \\
\hline
$S_K\supg$ & (Potential outcome) surrogate marker in Study $K (K=A,B)$ for group $g (g=0,1)$ \\
\hline
$Y_K\supg$ & (Potential outcome) primary outcome in Study $K (K=A,B)$ for group $g (g=0,1)$ \\
\hline
$S_{Kgi}$ & Observed surrogate marker for subject i, in group $g (g=0,1)$, in Study $K (K=A,B)$  \\
\hline
$Y_{K0i}$ & Observed primary outcome for subject i, in group $g (g=0,1)$, in Study $K (K=A,B)$; unobserved in Study B\\
\hline
$\mu_{Kg}(s)$ & $E(Y\supg_{K}| S\supg_{K}=s)$, the true conditional mean functions in Study K\\
\hline
$\Delta_B$ &  $E(Y_B\supone - Y_B\supzero)$, the unobserved true treatment effect on the primary outcome in Study B \\
\hline
$(\mu_1^*, \mu_0^*)$ &  Possible conditional mean functions\\
\hline
$\Delta_B(\mu_1^*, \mu_0^*)$ & $E\{\mu_1^*(S^{(1)}_B)-\mu_0^*(S^{(0)}_B)\}$, treatment effect in Study B given $(\mu_1^*, \mu_0^*)$\\
\hline
\multicolumn{2}{|l|}{\textbf{Resilience measures}} \\
\hline
Quantity & Description \\
\hline
$p_0$ & True resilience probability; fundamental parameter of interest  \\
\hline
$q_\alpha$ & True resilience bound (quantile); fundamental parameter of interest  \\
\hline
$\Omega_q$ & True resilience set; set of $\bPi$ ensuring the resilience bound $\geq 0$ \\
\hline
$\widehat{p}$ & Plug-in estimator of $p_0$ based on observed data  \\
\hline
$\widehat{q}_\alpha$ & Plug-in estimator of $q_\alpha$ based on estimated distribution $\widehat{F}_{\bPi}$ \\
\hline
$\widehat{\Omega}_q$ & Estimated resilience set based on empirical distribution $\widehat{F}_{\bPi}$ \\
\hline
\end{tabular}
\end{table}

\clearpage
\begin{table}[ht]
\centering
\renewcommand{\arraystretch}{1.1} 
\caption{Simulation results estimating the resilience probability, $p_0$, and resilience bound $q_{\alpha}$ for $\alpha=0.10$ in Settings 1-9 via the Gaussian Process (GP) algorithm (in Settings 1-3), the polynomial algorithm (Settings 4-6), and the Fourier series algorithm (Settings 7-9) in terms of bias, empirical standard error (ESE), and average standard error (ASE) estimated via the bootstrap. Note that for each setting, the algorithm used for estimation corresponded to the data generation (e.g. Gaussian Process algorithm was used for estimation in Settings 1-3); in Appendix C we additionally investigate robustness to misspecification of the basis class in these settings.\\}
\label{tab:gp_fixed_p}
\begin{tabular}{|c|l|c|c|c|c|c|}
\hline
\multicolumn{7}{|l|}{\textbf{Resilience probability, }$\mathbf{p_0}$} \\
\hline
Setting & Description &Truth &Estimate & Bias & ESE & ASE \\
\hline
1 & paradox likely, GP & 0.517 & 0.522 & -0.005 & 0.093 & 0.091 \\ \hline
2 & paradox possible, GP & 0.374 & 0.387 & -0.013 & 0.096 & 0.095 \\ \hline
3 & paradox unlikely, GP & 0.009 & 0.008 & 0.001 & 0.007 & 0.008 \\ \hline
4 & paradox likely, polynomial & 0.502 & 0.496 & 0.006 & 0.060 & 0.062 \\ \hline
5 & paradox possible, polynomial & 0.312 & 0.325 & -0.013 & 0.067 & 0.067 \\ \hline
6 & paradox unlikely, polynomial & 0.031 & 0.033 & -0.002 & 0.024 & 0.024 \\ \hline
7 & paradox likely, Fourier & 0.478 & 0.465 & 0.013 & 0.025 & 0.024 \\ \hline
8 & paradox possible, Fourier & 0.103 & 0.104 & -0.001 & 0.027 & 0.025 \\ \hline
9 & paradox unlikely, Fourier & 0.014 & 0.010 & 0.004 & 0.006 & 0.005 \\ \hline
\multicolumn{7}{|l|}{\textbf{Resilience bound, }$\mathbf{q_{\alpha}}$} \\
\hline
Setting & Description &Truth &Estimate & Bias & ESE & ASE \\
\hline
1 & paradox likely, GP & -1.058 & -1.017 & -0.042 & 0.184 & 0.187 \\ \hline
2 & paradox possible, GP & -0.723 & -0.682 & -0.042 & 0.183 & 0.185 \\ \hline
3 & paradox unlikely, GP & 1.197 & 1.285 & -0.087 & 0.263 & 0.272 \\ \hline
4 & paradox likely, polynomial & -2.911 & -2.969 & 0.058 & 0.544 & 0.509 \\ \hline
5 & paradox possible, polynomial & -1.598 & -1.617 & 0.019 & 0.486 & 0.464 \\ \hline
6 & paradox unlikely, polynomial & 0.918 & 0.865 & 0.053 & 0.393 & 0.391 \\ \hline
7 & paradox likely, Fourier & -1.950 & -1.976 & 0.025 & 0.167 & 0.133 \\ \hline
8 & paradox possible, Fourier & -0.008 & -0.004 & -0.004 & 0.074 & 0.068 \\ \hline
9 & paradox unlikely, Fourier & 0.500 & 0.524 & -0.024 & 0.068 & 0.055 \\ \hline
\end{tabular}
\end{table}

\clearpage

\clearpage

\clearpage
 \appendix

\setcounter{table}{0}
\setcounter{table}{0}
\renewcommand{\thetable}{A\arabic{table}}
\setcounter{figure}{0}
\renewcommand{\thefigure}{A\arabic{figure}}
\renewcommand{\theequation}{A.\arabic{equation}}

\clearpage

\section*{Appendix}

\section{Resilience Set Derivation}
In this section, we derive closed-form expressions for $F_{\bPi}$ for each function class considered. Once we have such an expression, the quantity $P(\Delta_B(\mu_1^*, \mu_0^*) < 0)$ can be expressed directly in terms of the parameters in $\bPi$, and can be used to identify regions of the parameter space where $q_\alpha \geq 0$. We begin with the  Gaussian Process class where we will show that $ \Delta_B(\mu_1^*, \mu_0^*) $ is approximately normally distributed, with mean and variance given as explicit functions. Let $\bS_{Bg}$ indicate the vector of $S_{Bgi}$ for $i = 1,...,n_{Bg}$ for $g=0,1$. and
$$\mathbf{m}_g = \begin{bmatrix} \widehat{\mu}_g(S_{g1}) \\ \vdots \\ \widehat{\mu}_g(S_{gn_g}) \end{bmatrix}, \quad
\bK_g = K_g(\bS_{Bg}, \bS_{Bg}; \sigma^2, \theta),$$
where $\bK_g$ is the $n_g \times n_g$ covariance matrix for group $g \in \{0,1\}$. Using straightforward properties of normal distributions, it follows that 
 $\Delta_B(\mu_1^*, \mu_0^*)$ is normally distributed i.e., 
$$ \Delta_B(\mu_1^*, \mu_0^*) \sim N(\mu_B, \sigma_B^2),$$
with
\begin{align*}
\mu_B &= \frac{1}{n_1} \mathbf{1}_{n_1}^\top \mathbf{m}_1 - \frac{1}{n_0} \mathbf{1}_{n_0}^\top \mathbf{m}_0, \\
\sigma_B^2 &= \frac{1}{n_1^2} \mathbf{1}_{n_1}^\top \bK_1 \mathbf{1}_{n_1} + \frac{1}{n_0^2} \mathbf{1}_{n_0}^\top \bK_0 \mathbf{1}_{n_0}.
\end{align*}

\noindent Thus, the distribution function $F_{\bPi}$ is
$$F_{\bPi}(z) = \Phi\left( \frac{z - \mu_B}{\sigma_B} \right),$$
\noindent and the resilience bound is given $\mu_B + \sigma_B \Phi^{-1}(\alpha),$ with resilience set $\left\{ \bPi \mid \mu_B + \sigma_B \Phi^{-1}(\alpha) \geq 0 \right\}.$
Note that since $\mathbf{m}_g$ and $\bK_g$ are defined based on the Study A and B data, it is feasible to solve for the parameters of $\bPi$ that belong to this set, as described in the main text. 
 
We now derive this distribution for the polynomial and Fourier series classes where $\bPi = \bSigma$. Define the matrices $\bM_g, g=0,1$ as the matrix of basis expansions on the $\bS_g$ values, so that 
\[
\bM_g = 
\begin{bmatrix}
1 & f_1(S_{Bg1}) & f_2(S_{Bg1}) & \cdots & f_{d-1}(S_{Bg1}) \\
1 & f_1(S_{Bg2}) & f_2(S_{Bg2}) & \cdots & f_{d-1}(S_{Bg2}) \\
\vdots & \vdots & \vdots & & \vdots \\
1 & f_1(S_{Bgn_g}) & f_2(S_{Bgn_g}) & \cdots & f_{d-1}(S_{Bgn_g})
\end{bmatrix},
\]
where the functions $f_1, \dots, f_{d-1}$ represent a prespecified basis: for the polynomial class, these are powers of standardized $s$ (e.g., $s, s^2, \dots$), and for the Fourier series class, these are the sine and cosine functions. Let $\bbeta = (\bbeta_0, \bbeta_1)^\top,$ where $\bbeta_0 = (\beta_{0,0}, \dots, \beta_{0,d-1})^\top$ and $\bbeta_1 = (\beta_{1,0}, \dots, \beta_{1,d-1})^\top$ denote the random coefficients, and 
$$\bbeta_0 \sim \mathcal{N}(0, \bSigma), \quad \bbeta_1 \sim \mathcal{N}(0, \bSigma),$$
with $\Sigma$ a diagonal $d \times d$ matrix. We can write the generated $\widehat{\by}_{Bg}$ as:
$$\begin{bmatrix}
\widehat{y}_{Bg1} \\
\vdots \\
\widehat{y}_{Bgn_g}
\end{bmatrix}
=
\begin{bmatrix}
\widehat{\mu}_g(S_{Bg1}) \\
\vdots \\
\widehat{\mu}_g(S_{Bgn_g}) 
\end{bmatrix}
+
\bM_g \bbeta_g.$$
Using straightforward properties of normal distributions:
$$
\frac{1}{n_0} \mathbf{1}^\top \widehat{\by}_{B0} \sim N \left( \frac{1}{n_0} \mathbf{1}^\top \widehat{\mu}_0(\bS_{B0}), \; \frac{1}{n_0^2} \mathbf{1}^\top \bM_0 \bSigma \bM_0^\top \mathbf{1} \right),
$$
$$
\frac{1}{n_1} \mathbf{1}^\top \widehat{\by}_{B1} \sim N \left( \frac{1}{n_1} \mathbf{1}^\top \widehat{\mu}_1(\bS_{B1}), \; \frac{1}{n_1^2} \mathbf{1}^\top \bM_1 \bSigma \bM_1^\top \mathbf{1} \right).$$

\noindent It follows that,
$$ \Delta_B(\mu_1^*, \mu_0^*) \sim N(\mu_B, \sigma_B^2),$$
with
\begin{align*}
\mu_B &= \frac{1}{n_1} \mathbf{1}^\top \widehat{\mu}_1(\bS_{B1}) - \frac{1}{n_0} \mathbf{1}^\top \widehat{\mu}_0(\bS_{B0}), \\
\sigma_B^2 &= \frac{1}{n_1^2} \mathbf{1}^\top \bM_1 \bSigma \bM_1^\top \mathbf{1} + \frac{1}{n_0^2} \mathbf{1}^\top \bM_0 \bSigma \bM_0^\top \mathbf{1}
\end{align*}

\noindent Thus, the distribution function $F_{\bPi}$ is
$$F_{\bPi}(z) = \Phi\left( \frac{z - \mu_B}{\sigma_B} \right),$$
\noindent and the resilience bound is given $\mu_B + \sigma_B \Phi^{-1}(\alpha),$ with resilience set $\left\{ \bPi \mid \mu_B + \sigma_B \Phi^{-1}(\alpha) \geq 0 \right\}.$
Similar to the Gaussian Process class, since $\widehat{\mu}_g(\bS_{Bg})$ and $\bM_g$ are defined based on the Study A and B data, it is theoretically possible to solve for the parameters of $\bPi$ that belong to this set, but unlike the Gaussian Process, $\bPi$ may contain more parameters depending on $d$ and assumptions about the structure of $\bSigma$ and thus, solving for this set in practice can be difficult.

\section{Theoretical Details}
\allowdisplaybreaks

\subsection{Unified Basis Representation}
We assume that the true conditional mean functions in Study B can be represented as a finite basis expansion:
\begin{equation}
\mu_{Bg}(s) = \sum_{k=1}^K W_{gk} C_k(s),\label{basis_1}
\end{equation}
for $g \in \{0,1\}$, where $\{C_k(\cdot)\}_{k=1}^K$ are known bounded basis functions, and $\{W_{gk}\}$ are random coefficients with a joint distribution governed by a finite set of parameters $\bPi \in \mathcal{P}$, with continuous density and finite second moments. We now show that each of our three proposed algorithms fit within this representation. We begin with the polynomial and Fourier series classes, as these are straightforward, and end with the Gaussian Process class, which is less straightforward. 
For all algorithms, the associated class of functions from which we draw $(\mu_1^*, \mu_0^*)$ is characterized as deviations from $\widehat{\mu}_0(s)$ and $\widehat{\mu}_1(s)$; therefore, we we will rewrite (\ref{basis_1}) as 
\begin{equation}
\mu_{Bg}(s) = \eta_g(s) + \sum_{k=2}^K W_{gk}C_k(s),\label{basis_2}
\end{equation}
where $W_{g1}=1$ and $C_1(s) = \eta_g(s)$, is a smooth function estimated nonparametrically from Study A as $\widehat{\mu}_g(s)$. For the polynomial class algorithm, $\mu_g^*(s)$ for $g=0,1$ are generated assuming that: 
$$\mu_{Bg}(s) =  \eta_0(s) + \sum_{j=0}^{d-1} \beta_{g,j} \cdot \left(\frac{s - \bar{s}_{Bg}}{z_{Bg}} \right)^j$$
which is of the unified basis form with $W_{gk} = \beta_{g,k-2}$ and $C_k(s) = \left(\frac{s - \bar{s}_{Bg}}{z_{Bg}} \right)^{k-2}$ and $K=d-1$. For the Fourier series class algorithm $\mu_g^*(s)$ for $g=0,1$ are generated assuming that: 
$$\mu_{Bg}(s) =  \eta_g(s) + \beta_{g,0} + \sum_{j=1}^{d-1} \beta_{g, j} \left[\sin\left(\frac{s - c_g}{B_j}\right) + cos\left(\frac{s - c_g}{B_j}\right)\right]$$
which is of the unified basis form with $W_{g2} = \beta_{g,0}$ and $W_{gk} = \beta_{g,k-2}$ for $k=3,...,K$; $C_2(s) = 1$ and $C_k(s) = \left[\sin\left(\frac{s - c_g}{B_{k-2}}\right) + cos\left(\frac{s - c_g}{B_{k-2}}\right)\right]$ and $K=d-1$.

For the Gaussian Process algorithm, we use the Karhunen–Loève expansion \citep{adler2009random}, which states that a stochastic process can be represented as an infinite linear combination of orthogonal functions, to represent the process as:
$$\mu_{Bg}(s) = \eta_g(s) + \sum_{k=2}^\infty W_{gk} \phi_k^{(g)}(s)$$
where $\phi_k^{(g)}(s)$ are orthonormal eigenfunctions of the RBF kernel and $W_{gk} \sim \mathcal{N}(0, \lambda_k^{(g)})$ are independent random coefficients with variances equal to the corresponding eigenvalues. Truncating to $K$ terms, which we justify below, yields:
$$\mu_{Bg}(s) \approx \eta_g(s) + \sum_{k=2}^K W_{gk} \phi_k^{(g)}(s),$$
which is of the unified basis form with $C_k(s) = \phi_k^{(g)}(s)$.

It remains to show that our truncation to $K$ terms above is asymptotically appropriate. To do so, we need to justify that this finite approximation is close enough for both 1) function approximation and 2) our downstream quantities derived from the function in our proposed procedure which include empirical averages and the distribution of these empirical averages. We begin with showing function approximation. Let

$$\tau(s) = \sum_{k=2}^\infty \sqrt{\lambda_k} W_k C_k(s), \quad \text{and} \quad \tau_{K}(s) = \sum_{k=2}^{K} \sqrt{\lambda_k} W_k C_k(s),$$
where, for clarity, we focus solely on the summations only (dropping the first term above) and drop the $g$ from our notation, and take $W_k \sim \mathcal{N}(0,1)$ i.i.d. and $C_k(s)$ are orthonormal basis functions from the expansion described above.  Here $\{\lambda_n\}$ are eigenvalues corresponding to the specific kernel and we assume that $\sum_{i=1}^\infty \lambda_i<\infty.$  To show that $\tau_K(s)$ approximates $\tau(s)$ in $L^2$, consider the integrated mean squared error:
\begin{equation}
E \left[ \int_{-\infty}^\infty \left\{ \tau(s) - \tau_K(s) \right\}^2 ds \right] = \sum_{k=K+1}^\infty \lambda_k,
\end{equation}
which follows from orthogonality and independence. Since $\sum_{i=1}^\infty \lambda_i<\infty$, the eigenvalues $\lambda_k$ decay sufficiently fast and this $L^2$ error becomes negligible as $K \rightarrow \infty.$ This tells us that the truncated process approximates the full process in functional space. 

While this controls errors in functional approximation, our proposed procedure requires investigating the asymptotic distribution of the empirical averages over samples, such as:
$$
\frac{1}{\sqrt{n}} \sum_{i=1}^n \tau(S_i),
$$
where $S_i \sim F$ i.i.d. random variables. To this end, consider:
$$
\frac{1}{\sqrt{n}} \sum_{i=1}^n \tau(S_i) 
=  \frac{1}{\sqrt{n}} \sum_{i=1}^n \tau_K(S_i) + \frac{1}{\sqrt{n}} \sum_{i=1}^n \left[ \tau(S_i) - \tau_K(S_i) \right],
$$
where the second term is the residual that we wish to be negligible. By independence and using the same orthogonality argument:
\begin{align*}
E \left\{ \left(\frac{1}{\sqrt{n}}\sum_{i=1}^n\left[ \tau(S_i) - \tau_K(S_i) \right]\right)^2 \right\} =& \int_{-\infty}^{\infty} E\left\{\tau(s)-\tau_K(s)\right\}^2f_S(s)ds\\
\le & C_0\sum_{k=K+1}^\infty \lambda_k \to 0 \quad \text{as } K \to \infty,
\end{align*}
where $f_S(\cdot)$ is the density function of $S_i$ bounded from above by a positive constant $C_0.$
Hence, for any $\delta>0,$
\begin{equation}
P\left(\lim_{K\rightarrow 0} \biggm|\frac{1}{\sqrt{n}} \sum_{i=1}^n \left[ \tau(S_i) - \tau_K(S_i) \right]\biggm|>\delta \right) \rightarrow 0
\label{eq:diffbound}
\end{equation}

This shows that empirical averages computed with the truncated process are close in probability to those from the full process. Lastly, because our procedure relies on the distribution of the average, we will show that the truncated empirical average is approximately Gaussian. Define the centered feature vector
$$
\mathbf{Z}_i = \begin{pmatrix}
W_1C_1(S_i) \\
\vdots \\
W_KC_K(S_i)
\end{pmatrix}, 
$$
where $K \rightarrow \infty$ and $K/n\rightarrow 0.$  Under mild regularity conditions (e.g., sub-Gaussian tails and finite second moments), a high-dimensional central limit theorem (CLT) applies:
$$
\frac{1}{\sqrt{n}} \sum_{i=1}^n \mathbf{Z}_i \xrightarrow{d} N(0, \Sigma_K),
$$
for some positive definite matrix $\Sigma_K$ in the sense that 
$$ \mathbf{w}_K'  \frac{1}{\sqrt{n}}\sum_{i=1}^n \mathbf{Z}_i \xrightarrow{d} N(0, \mathbf{w}_K'\Sigma_K\mathbf{w}_K)$$
for any $K$ dimensional vector $\mathbf{w}_K.$  Let $\mathbf{w}_K=(\sqrt{\lambda_1}, \cdots, \sqrt{\lambda_K})',$ we may conclude that
$$
\frac{1}{\sqrt{n}} \sum_{i=1}^n \tau_K(S_i)  \xrightarrow{d} N(0, \sigma_K^2),$$
where 
$$\sigma_K^2\rightarrow \sigma_0^2=E\left\{k(S_1, S_1)\right\}+E\left\{k(S_1, S_2)\right\}$$ 
as $K \rightarrow \infty. $  
Now, for any $\epsilon>0,$ we may select a sufficiently large $K$ such that 
$$
P\left( \frac{1}{\sqrt{2\pi}\sigma_0}\biggm|\frac{1}{\sqrt{n}} \sum_{i=1}^n \left[ \tau(S_i) - \tau_K(S_i) \right]\biggm|>\epsilon \right) < \epsilon
$$
and
$$\sup_x|\Phi(x, \sigma_K)-\Phi(x, \sigma_0)|<\epsilon,$$
where $\Phi(x, \sigma)$ is the CDF of a Gaussian distribution with mean zero and variance $\sigma^2.$ With this selected $K,$ we choose $n$ sufficiently large, i.e., $n>N_0$, such that 
$$\sup_x \biggm| P\left(\frac{1}{\sqrt{n}}\sum_{i=1}^n \tau_K(S_i)\le x\right)- \Phi(x; \sigma_K)  \biggm| < \epsilon.$$
Therefore, 
\begin{align*}
& \sup_x \biggm| P\left(\frac{1}{\sqrt{n}}\sum_{i=1}^n \tau(S_i)\le x\right)- \Phi(x; \sigma_0)  \biggm| \\
\le & \sup_x \biggm| P\left(\frac{1}{\sqrt{n}}\sum_{i=1}^n \tau_K(S_i)\le x-\frac{1}{\sqrt{n}}\sum_{i=1}^n [\tau(S_i)-\tau_K(S_i)]\right)- \Phi\left(x-\frac{1}{\sqrt{n}}\sum_{i=1}^n [\tau(S_i)-\tau_K(S_i)]; \sigma_K \right)  \biggm| \\
&+\sup_x \biggm|\Phi\left(x-\frac{1}{\sqrt{n}}\sum_{i=1}^n [\tau(S_i)-\tau_K(S_i)]; \sigma_K \right)- \Phi\left(x-\frac{1}{\sqrt{n}}\sum_{i=1}^n [\tau(S_i)-\tau_K(S_i)]; \sigma_0 \right) \biggm|\\
&+\sup_x \biggm|P\left( \sigma_0 W\le x-\frac{1}{\sqrt{n}}\sum_{i=1}^n [\tau(S_i)-\tau_K(S_i)] \right)- P\left(\sigma_0W\le x\right) \biggm|\\
\le &\epsilon+\epsilon+ P\left(\frac{1}{\sqrt{2\pi}\sigma_0}\biggm|\frac{1}{\sqrt{n}}\sum_{i=1}^n [\tau(S_i)-\tau_K(S_i)]\biggm|>\epsilon\right)\\
&+\sup_x \biggm|P\left( \sigma_0 W\le x-\frac{1}{\sqrt{n}}\sum_{i=1}^n [\tau(S_i)-\tau_K(S_i)],  \frac{1}{\sqrt{2\pi}\sigma_0}\biggm|\frac{1}{\sqrt{n}}\sum_{i=1}^n [\tau(S_i)-\tau_K(S_i)]\biggm|<\epsilon \right)- P\left(\sigma_0W\le x\right) \biggm|\\
\le & 4\epsilon
\end{align*}
for arbitrarily small $\epsilon>0,$ where $W\sim N(0, 1)$ is a standard normal random variable independent of data.  Therefore,
$$
\frac{1}{\sqrt{n}} \sum_{i=1}^n \tau(S_i) \xrightarrow{d} N(0, \sigma_0^2),$$
in distribution, as $n \rightarrow \infty.$  In summary, this justifies our use of a finite basis expansion in our framework. Even though we truncate the process, the sample-level quantities we care about (e.g., averages) behave as if they came from the full process, for large enough $K$.

\subsection{Proof of Theorem 1}

Under our stated assumptions, we first express $p_0$ as a functional of the true CDF $F_{S_B^{(g)}}(\cdot),$ $g=0, 1$, as follows:
\begin{align*}
 p_0=&P\left(\Delta_B(\mu_1^*, \mu_0^*)<0\right)\\
 =&P\left(\int \mu_1^*(s)dF_{S_B^{(1)}}(s)-\int \mu_0^*(s)dF_{S_B^{(0)}}(s)<0\right)\\
 =&P\left( \sum_{k=1}^K W_{1k} \int C_k(s)dF_{S_B^{(1)}}(s)- \sum_{k=1}^K W_{0k}\int C_k(s)dF_{S_B^{(0)}}(s)<0\right) \\
 =& h\left\{\left(\int C_1(s)dF_{S_B^{(1)}}(s), \cdots, \int C_K(s)dF_{S_B^{(1)}}(s), \int C_1(s)dF_{S_B^{(0)}}(s), \cdots, \int C_K(s)dF_{S_B^{(0)}}(s)\right), 0\right\}\\
 =& h(\theta, 0)
 \end{align*}
where $h(\cdot)$ is a differentiable function and
$$\theta = \left ( \int C_1(s)dF_{S_B^{(1)}}(s), \cdots, \int C_K(s)dF_{S_B^{(1)}}(s), \int C_1(s)dF_{S_B^{(0)}}(s), \cdots, \int C_K(s)dF_{S_B^{(0)}}(s) \right ),$$ 
Let $\widehat{\theta}$ denote the corresponding plug-in estimator with $F_{S_B^{(g)}}(\cdot), j=0, 1$ replaced with the empirical CDF $\widehat{F}_{S_B^{(g)}}(\cdot)$. Since $\widehat{p}=h(\widehat{\theta}, 0)$
and $\widehat{\theta}  \xrightarrow{P} \theta$, it follows that 
$$\widehat{p} = h(\widehat{\theta}, 0)  \xrightarrow{P} h(\theta, 0) = p_0.$$
In addition,  
$$\sqrt{\frac{n_{B1}n_{B0}}{n_{B1}+n_{B0}}}\left\{\left(\begin{array}{c} \int C_1(s)dF_{S_B^{(1)}}(s) \\ \vdots \\ \int C_K(s)dF_{S_B^{(0)}}(s) \end{array} \right)-\left(\begin{array}{c} \int C_1(s)d\widehat{F}_{S_B^{(1)}}(s) \\ \vdots \\ \int C_K(s)d\widehat{F}_{S_B^{(0)}}(s) \end{array} \right)\right\} \rightarrow N(0, \Upsilon),$$
 for some asymptotic covariance matrix $\Upsilon$ that depends on the variability in the empirical CDFs and the functions $\{C_k\}$. Applying the multivariate delta method to the differentiable function $h$, since $n_{B0}/(n_{B0}+n_{B1})=r \in (0, 1),$ it follows that  
$$\sqrt{\frac{n_{B1}n_{B0}}{n_{B1}+n_{B0}}}(\widehat{p}-p_0) \rightarrow N(0, \sigma_P^2)$$
where 
$$\sigma_P^2 = \frac{d h(\theta, 0)}{d \theta}^\top \Upsilon \frac{d h(\theta, 0)}{d \theta}.$$

\subsection{Proof of Theorem 2}

We first establish consistency of $\widehat{q}_\alpha$. Noting that
\begin{align*}
 F_{\bPi}(\delta)=&P\left(\Delta_B(\mu_1^*, \mu_0^*)<\delta\right)\\
 =&P\left(\int \mu_1^*(s)dF_{S_B^{(1)}}(s)-\int \mu_0^*(s)dF_{S_B^{(0)}}(s)<\delta\right)\\
 =&P\left( \sum_{k=1}^K W_{1k} \int C_k(s)dF_{S_B^{(1)}}(s)- \sum_{k=1}^K W_{0k}\int C_k(s)dF_{S_B^{(0)}}(s)<\delta\right) \\
 =& h(\theta, \delta)
 \end{align*}
$\widehat{F}_{\bPi}(\delta)=h(\widehat{\theta}, \delta).$ Coupled with the fact that $\widehat{\theta}=\theta+o_p(1),$ it implies that $\widehat{F}_{\bPi}$ converges uniformly to $F_{\bPi}$
$$\sup_{\delta} |\widehat{F}_{\bPi}(\delta) - F_{\bPi}(\delta)| =o_p(1).$$
By the regularity condition, $q_\alpha$ is the unique solution to $F_{\bPi}(\delta)=\alpha$ and $f_{\bPi}(q_\alpha)\neq 0,$ it follows that $\widehat{q}_\alpha \xrightarrow{p} q_\alpha$. 
Next, we have 
\begin{align*}
0=&\widehat{F}_{\bPi}(\widehat{q}_\alpha)-F_{\bPi}(q_\alpha)\\
 =&\widehat{F}_{\bPi}(\widehat{q}_\alpha)-\widehat{F}_{\bPi}(q_\alpha)+\widehat{F}_{\bPi}(q_\alpha)-F_{\bPi}(q_\alpha)\\
 =& \frac{\partial h(\theta, q_\alpha)}{\partial q}(\widehat{q}_\alpha-q_\alpha)+\frac{\partial h(\theta, q_\alpha)}{\partial \theta}^\top(\widehat{\theta}-\theta)+o\left(|\widehat{q}_\alpha -q_\alpha|+\left(\frac{n_{B1}n_{B0}}{n_{B1}+n_{B0}}\right)^{-1/2}\right)
\end{align*}
which implies that
$$  \sqrt{\frac{n_{B1}n_{B0}}{n_{B1}+n_{B0}}}(\widehat{q}_\alpha - q_\alpha)=\left(\frac{\partial h(\theta, q_\alpha)}{\partial q} \right)^{-1}\frac{\partial h(\theta, q_\alpha)}{\partial \theta}^\top \sqrt{\frac{n_{B1}n_{B0}}{n_{B1}+n_{B0}}} (\widehat{\theta}-\theta)+o(1).$$
Noting that 
$$ \frac{\partial h(\theta, q_\alpha)}{\partial q} =f_{\bPi}(q_\alpha)$$ is the value of the density function associated with $F_{\bPi}$ at $q_\alpha,$ 
$$\sqrt{\frac{n_{B1}n_{B0}}{n_{B1}+n_{B0}}}(\widehat{q}_\alpha - q_\alpha) \xrightarrow{d} N(0, \sigma_Q^2),$$
where
$$\sigma_Q^2=\frac{ \frac{\partial h(\theta, q_\alpha)}{\partial \theta}^\top \Upsilon \frac{\partial h(\theta, q_\alpha)}{\partial \theta}}{[f_{\bPi}(q_\alpha)]^2}.$$

\subsection{Proof of Theorem 3}
By Assumption 6, $\widehat{q}_\alpha(\widehat{F}_{\bPi})$ converges to $q_\alpha(F_{\bPi})$ uniformly in $\bPi \in \mathcal{P}.$ Coupled with the continuity of the map $\bPi \mapsto q_\alpha(F_{\bPi})$, it implies that for any $\epsilon > 0$, there exists $N$ such that for all $n_B > N$,

$$\sup_{\bPi \in \mathcal{P}} |\widehat{q}_\alpha(\widehat{F}_{\bPi}) - q_\alpha(F_{\bPi})| < \epsilon$$
with probability at least $1 - \delta$. Since $\Omega_q = \{\bPi \in \mathcal{P} : q_\alpha(F_{\bPi}) \geq 0\}$ and $\widehat{\Omega}_q = \{\bPi \in \mathcal{P} : \widehat{q}_\alpha(\widehat{F}_{\bPi}) \geq 0\}$, the sets become arbitrarily close in Hausdorff distance. That is, $d_H(\widehat{\Omega}_q, \Omega_q) \xrightarrow{p} 0$. This follows as a standard consequence of uniform convergence of estimators and continuity of the underlying functional over a compact parameter space; see, e.g., \citet[Section 5.9]{van2000asymptotic}.

\section{Details and Additional Results: Simulations and Data Application}
\allowdisplaybreaks

\subsection{Simulation Details: Data Generation}
\label{sim-details}
\begin{figure}[ht]
    \centering
     \vspace*{-10mm}
    \includegraphics[width=1\textwidth]{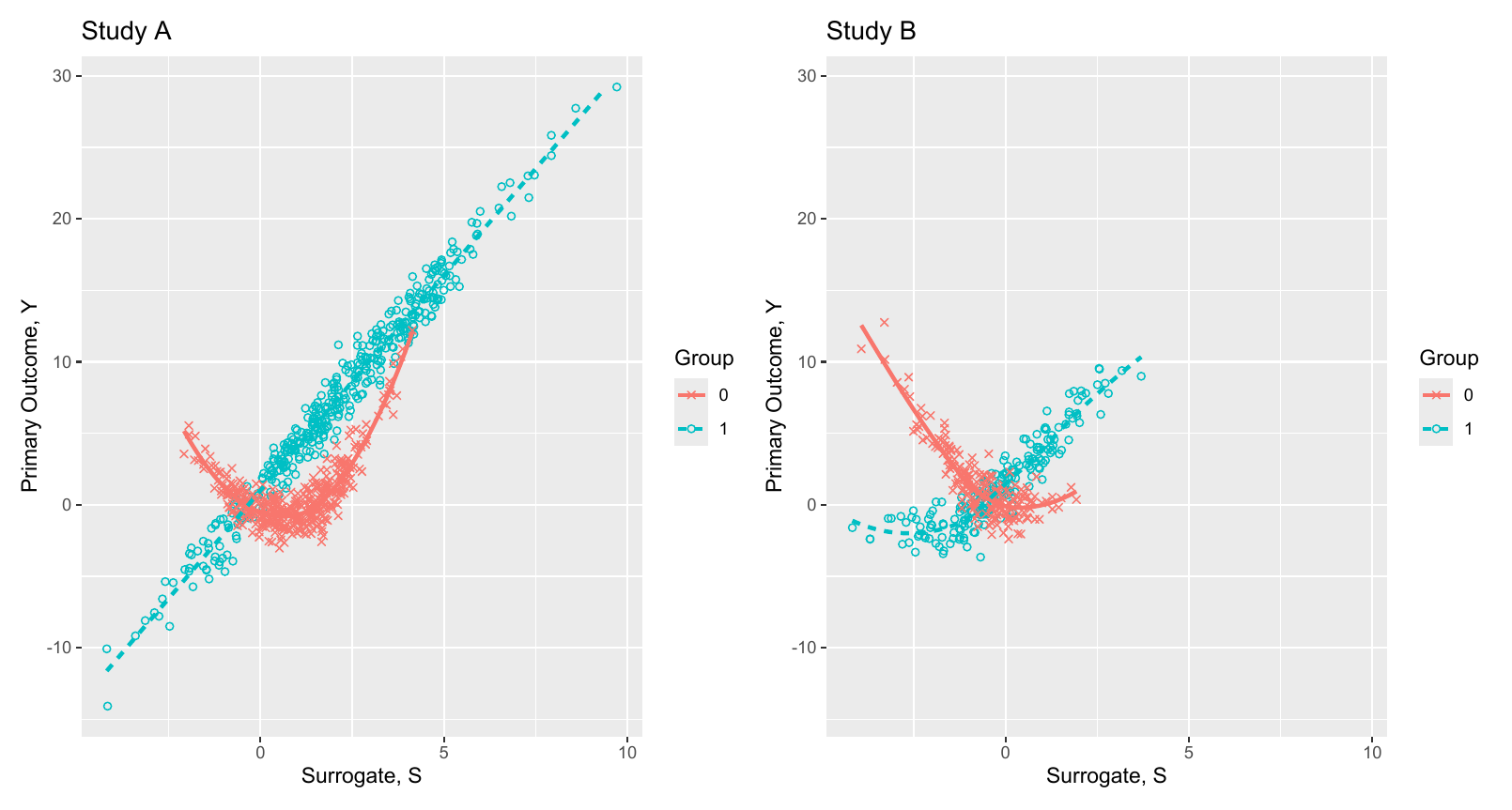}
    \caption{Single iteration of Setting 4 displaying $Y$ as a function of $S$ for each treatment group in Study A (left panel) and Study B (right panel). Note that in practice, we would not be able to construct the Study B figure as it requires $Y$ in Study B.}
    \label{fig_setting4}

\end{figure}

Here, we provide details of the data generation in each simulation setting, as well as motivation for these settings. We begin with discussing the motivation. Clearly, our proposed methods rely on the basis function specifications described in the main text. Thus, at a minimum we needed to generate one setting assuming a Gaussian Process truth, one setting assuming a polynomial function truth, and one setting assuming a Fourier series truth. However, we also wanted to examine how our methods perform when the surrogate paradox is likely versus unlikely. That is, can we correctly estimate a large $p_0$ when it is large, and a small $p_0$ when it is small? This motivated us to consider three ``tiers", i.e. settings where the paradox was probable vs. possible vs. unlikely in Study B. In addition, for all settings, we need 1) the paradox to \textit{not} be present in Study A (because then the entire exercise would not be carried out), 2) the surrogate to be a reasonably strong surrogate in Study A (again, if this was not the case, the entire exercise would not be carried out; we defined this as the surrogate having an estimated proportion of treatment effect explained quantity of 0.50 or larger), and 3) the surrogate in Study B to reflect a treatment effect on $S$ (once again, if this was not the case, the entire exercise would not be carried out). We will now walk through a single iteration of Setting 4 specifically, as an illustration. In the left panel of Figure \ref{fig_setting4}, we show a single iteration of generated data for each group from Study A with a smoothed fitted curve. In this dataset, there is a positive treatment effect on $Y$, a positive treatment effect on $S$, and the estimated proportion of the treatment effect explained by the surrogate is 0.54. In the right panel of Figure \ref{fig_setting4}, we show a single iteration of generated data for each group from Study B with a smoothed fitted curve. Note that in practice, we would not be able to create this plot because we would not have the primary outcome, $Y$, in Study B. Of course, as is the theme of the paper, the data were generated from a conditional mean function that is not exactly equal to the conditional mean function in Study A. To induce a high probability of the surrogate paradox occurring within Study B for this setting, we also purposefully generate the \textit{surrogates} for Study B to come from the ``problematic" region of Study A where the two curves cross. This is highlighted by keeping the axes the same across the two plots. In this specific Study B, there is a positive treatment effect on the surrogate marker, but there is a negative treatment effect on the primary outcome. In this setting, because we purposefully generated the data in this way in Study B, the true probability of the surrogate paradox occurring in Study B is 0.502. We can make this probability smaller by generating the surrogate in Study B further from the area where the curves cross; this is precisely how we generate data for Setting 5 (true probability is 0.312) and Setting 6 (true probability is 0.031). Below, we provide the full data generation details for all settings.

We investigated nine simulation settings. In Settings 1-3, the data were generated such that the Gaussian process function specification was correct. In Settings 4-6, the data were generated such that the polynomial function specification was correct. In Settings 7-9, the data were generated such that the Fourier series function specification was correct. In Settings 1, 4, and 7, the occurrence of the surrogate paradox in Study B was probable: $p_0 = 0.517$ in Setting 1, $p_0 = 0.502$ in Setting 4, and $p_0 = 0.478$ in Setting 7. In Settings 2, 5, and 8, the occurrence of the surrogate paradox in Study B was possible: $p_0 = 0.374$ in Setting 2, $p_0 = 0.312$ in Setting 5, and $p_0 = 0.103$ in Setting 8. In Settings 3, 6, and 9, the occurrence of the surrogate paradox in Study B was unlikely: $p_0 = 0.009$ in Setting 3, $p_0 = 0.031$ in Setting 6, and $p_0 = 0.014$ in Setting 9.

First, for each setting, we give explicit details below for a) the specified distributions of the surrogate in each group, in each study, b) the true conditional mean functions in each group e.g. $m_0(s)$ and $m_1(s)$, and c) defined parameters to be used in data generation, including variance parameters. (Throughout, normal distributions are given as $N(\mu, \sigma^2)$ where $\mu$ denotes the mean and $\sigma^2$ denotes the variance.)

\noindent \textbf{Setting 1} \\
$S_A\supzero \sim N(3,3)$, $S_A\supone \sim N(4,3)$, $S_B\supzero \sim N(4.75,1)$, $S_B\supone \sim N(5.25, 1)$, $m_0(s) = 2s - 1$, $m_1(s) = s + 3$, $\sigma^2 = 0.30$, $\theta = 5$, $v_0^2 = v_1^2 = 1$ \\
\textbf{Setting 2} \\
$S_A\supzero \sim N(3,3)$, $S_A\supone \sim N(4,3)$, $S_B\supzero \sim N(4.75,1)$, $S_B\supone \sim N(5.25, 1)$, $m_0(s) = 2s - 1.25$, $m_1(s) = s + 3$, $\sigma^2 = 0.25$, $\theta = 5$, $v_0^2 = v_1^2 = 1$ \\
\textbf{Setting 3}\\
$S_A\supzero \sim N(2,3)$, $S_A\supone \sim N(3,3)$, $S_B\supzero \sim N(1.75,1)$, $S_B\supone \sim N(2.75, 1)$, $m_0(s) = 1.5s + 1$, $m_1(s) = 3s-2$, $\sigma^2 = 1$, $\theta = 1$, $v_0^2 = v_1^2 = 3$ \\
\textbf{Setting 4} \\
$S_A\supzero \sim N(0.9,1.5)$, $S_A\supone \sim N(2.2,4.5)$, $S_B\supzero \sim N(-0.7,1)$, $S_B\supone \sim N(-0.2, 2)$, $m_0(s) = (s - 0.5)^2-1$, $m_1(s) = 3s + 1$, $\Sigma = diag(0.25, 0.25, 0.1, 0.1)$, $v_0^2 = v_1^2 = 1$ \\
\textbf{Setting 5} \\
$S_A\supzero \sim N(0.9,1.5)$, $S_A\supone \sim N(2.2,4.5)$, $S_B\supzero \sim N(-0.5,1)$, $S_B\supone \sim N(0, 2)$, $m_0(s) = (s - 0.5)^2-1$, $m_1(s) = 3s + 1$, $\Sigma = diag(0.25, 0.25, 0.1, 0.1)$, $v_0^2 = v_1^2 = 1$ \\
\textbf{Setting 6} \\
$S_A\supzero \sim N(0.9,1.5)$, $S_A\supone \sim N(2.2,4.5)$, $S_B\supzero \sim N(-0.08,1)$, $S_B\supone \sim N(0.45, 2)$, $m_0(s) = (s - 0.5)^2-1$, $m_1(s) = 3s + 1$, $\Sigma = diag(0.25, 0.25, 0.1, 0.1)$, $v_0^2 = v_1^2 = 1$ \\
\textbf{Setting 7} \\
$S_A\supzero \sim N(5,1)$, $S_A\supone \sim N(6,2)$, $S_B\supzero \sim N(4.1,0.5)$, $S_B\supone \sim N(4.5, 0.5)$, $m_0(s) = 0.2 + 0.4\sin(s) + 0.4\cos(s)$, $m_1(s) = 0.6 + 0.85\sin(s) + 0.85\cos(s)$, $\Sigma = diag(0.5, 0.5, 0.1, 0.1)$, $v_0^2 = v_1^2 = 0.05$ \\
\textbf{Setting 8} \\
$S_A\supzero \sim N(5,1)$, $S_A\supone \sim N(6,2)$, $S_B\supzero \sim N(4.7,1)$, $S_B\supone \sim N(5.4, 1)$, $m_0(s) = 0.2+0.4\sin(s)+0.4\cos(s)$, $m_1(s) = 0.6+ 0.85\sin(s)+0.85\cos(s)$, $\Sigma = diag(0.05, 0.05, 0.01, 0.01)$, $v_0^2 = v_1^2 = 0.05$ \\
\textbf{Setting 9} \\
$S_A\supzero \sim N(5,1)$, $S_A\supone \sim N(6,2)$, $S_B\supzero \sim N(5.5,0.5)$, $S_B\supone \sim N(6.5, 0.5)$, $m_0(s) = 0.2+0.4\sin(s)+0.4\cos(s)$, $m_1(s) = 0.6+ 0.85\sin(s)+0.85\cos(s)$, $\Sigma = diag(0.05, 0.05, 0.01, 0.01)$, $v_0^2 = v_1^2 = 0.05$ \\

\noindent Next, we specify the data distribution given these specifications. In Settings 1-3 where the 
GP data-generating process is true, the data are generated as: 
\begin{eqnarray*}
Y_A\supzero  &\sim& MVN(m_0(S_A\supzero), K(S_A\supzero, S_A\supzero; \sigma^2, \theta)) + \epsilon_{0A}\\ 
Y_A\supone  &\sim& MVN(m_1(S_A\supone), K(S_A\supone, S_A\supone; \sigma^2, \theta)) + \epsilon_{1A}\\
Y_B\supzero  &\sim& MVN(\widehat{\mu}_0(S_B\supzero), K(S_B\supzero, S_B\supzero; \sigma^2, \theta)) + \epsilon_{0B}\\ 
Y_B\supone &\sim& MVN(\widehat{\mu}_1(S_B\supone), K(S_B\supone, S_B\supone; \sigma^2, \theta)) + \epsilon_{1B} 
\end{eqnarray*}
where $\epsilon_{0A}, \epsilon_{0B}, \epsilon_{1A}, \epsilon_{1B}$ are each iid $N(0, v_0^2)$ and $K$ denotes the radial basis covariance kernel. In Settings 4-6 where the polynomial data-generation process is true, the data are  generated as:
$$Y_A\supzero = m_0(S_A\supzero) + \epsilon_{0A}$$
$$Y_A\supone = m_1(S_A\supone) + \epsilon_{1A}$$
$$Y_B\supzero = \widehat{\mu}_0(S_A\supzero) + \sum_{i=0}^{K-1} \beta_{0,i} \cdot \left (\frac{S_B\supzero - \bar{s}_{A0}}{z_{A0}} \right )^i + \epsilon_{0B}$$
$$Y_B\supone = \widehat{\mu}_1(S_A\supone) + \sum_{i=0}^{K-1} \beta_{1,i} \cdot \left (\frac{S_B\supone - \bar{s}_{A1}}{z_{A1}} \right )^i + \epsilon_{1B}$$
where $\bar{s}_{Ag}$ and $z_{Ag}$ are the sample mean and standard deviation, respectively, of $S_{Ag}$ used to standardize $s$, $\beta_0,\beta_1 \sim MVN(0, \Sigma)$, $K=4$, and $\epsilon_{0A}, \epsilon_{0B}, \epsilon_{1A}, \epsilon_{1B}$ are each iid $N(0, v_0^2)$.  In Settings 7-9 where the Fourier series data-generation process is true, the data are  generated as:
$$Y_A\supzero = m_0(S_A\supzero) + \epsilon_{0A}$$
$$Y_A\supone = m_1(S_A\supone) + \epsilon_{1A}$$
$$Y_B\supzero = \widehat{\mu}_0(S_A\supzero) + \sum_{j=1}^{d-1} \beta_{0, j} \left[\sin\left(\frac{S_{B}\supzero - c_0}{B_{0j}}\right) + \cos\left(\frac{S_{B}\supzero - c_0}{B_{0j}}\right)\right] + \epsilon_{0B}$$
$$Y_B\supone = \widehat{\mu}_1(S_A\supone) + \sum_{j=1}^{d-1} \beta_{1, j} \left[\sin\left(\frac{S_{B}\supone - c_1}{B_{1j}}\right) + \cos\left(\frac{S_{B}\supone - c_1}{B_{1j}}\right)\right] + \epsilon_{1B}$$
where $d=4$, $c_0 = \min(S_{A0i})$, $c_1 = \min(S_{A1i})$, and $\epsilon_{0A}, \epsilon_{0B}, \epsilon_{1A}, \epsilon_{1B}$ are each iid $N(0, v_0^2)$, $\beta_0,\beta_1 \sim MVN(0, \Sigma)$, and the constants $B_{gj}$ which control the period were set to $B_{g1} = 2\pi/(0.5r_{gA})$, $B_{g2} = 2\pi/(0.25r_{gA})$, and $B_{g3} = 2\pi/(0.1r_{gA})$ where $r_{gA}$ is the observed range of $S_{gA}$.
\\

\subsection{Simulations: Additional Results}

In this section we present results investigating robustness to misspecification of the basis class. Specifically, we took one setting from each set (Setting 1 from Settings 1-3; Setting 4 from Settings 4-6; Setting 9 from Settings 7-9) such that we also had one setting where the paradox was probable (Setting 4), possible (Setting 2), and unlikely (Setting 9). For each, we estimated the resilience probability and resilience bound using the incorrect algorithm. That is, for Setting 2 which was generated using the Gaussian Process specification, estimation was done using the Fourier series algorithm with $\Sigma = diag(0.25, 0.25, 0.1, 0.1)$, $B = (0.5,  0.25, 0.1)$. For Setting 4 which was generated using the polynomial specification, estimation was done using the Gaussian Process algorithm with $\sigma^2 = 0.25$, $\theta = 2$. For Setting 9 which was generated using the Fourier series specification, estimation was done using the polynomial algorithm with $\Sigma = diag(0.05, 0.05, 0.01, 0.01)$. Results are shown in Table \ref{tab:misspecified-simulations}. While we do see more bias here compared to when the correct algorithm is used in the main text, as expected, the bias for the resilience probability is reasonably small. The estimated resilience bounds are also biased but still directionally correct (when the truth is negative/positive, the estimate is also negative/positive).

\subsection{Data Application: Additional Results}

\begin{figure}[ht]
    \centering
    \includegraphics[width=0.8\textwidth]{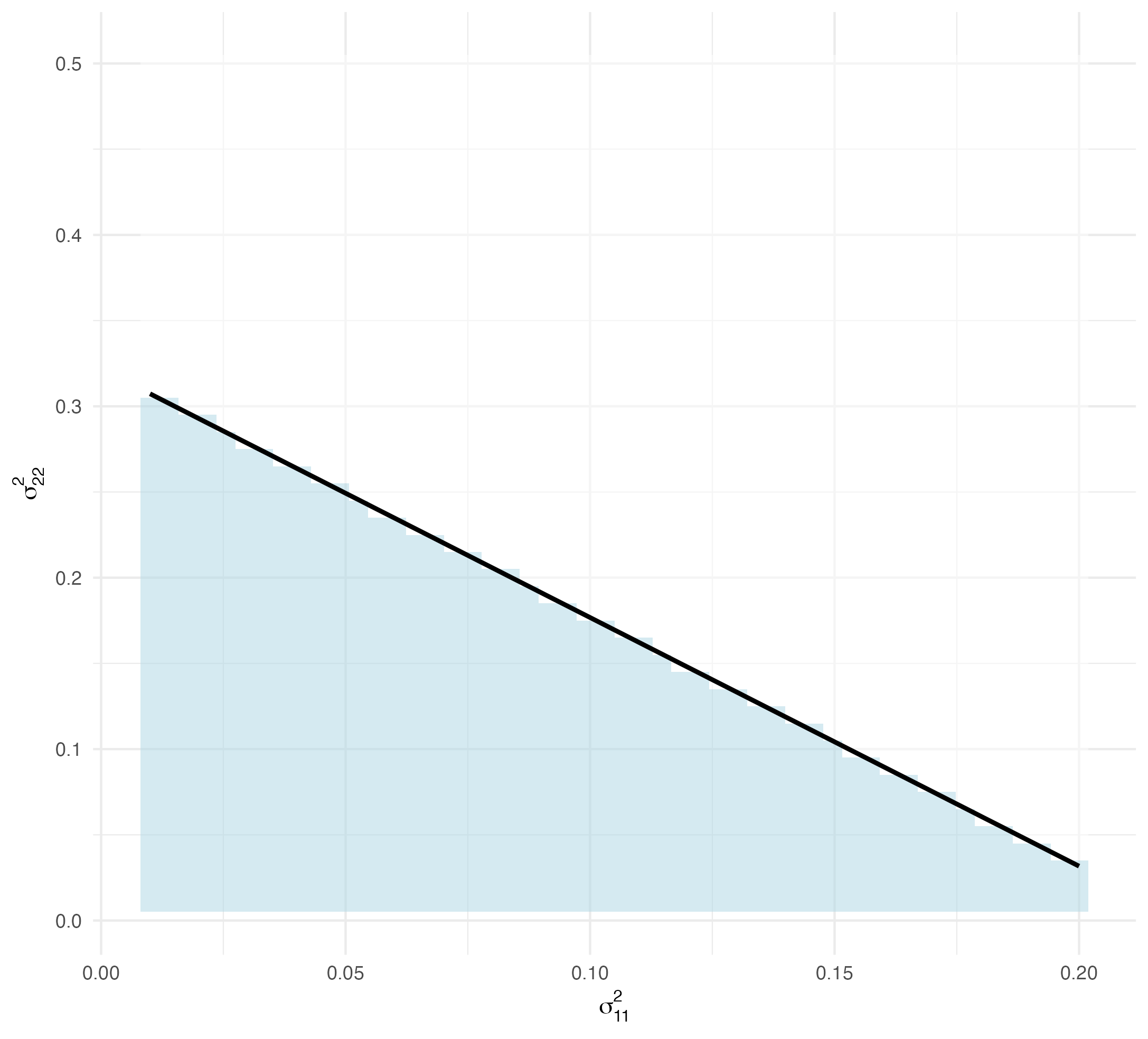}
    \caption{Estimated resilience set, $\widehat{\Omega}_q =  \{\bPi \mid \widehat{q}_\alpha(\widehat{F}_{\bPi}) \geq 0\}$, for the HIV data application using the polynomial algorithm parameterized by $\bSigma$, with an $\alpha=0.10$; blue shading indicates the identified pairings of $\sigma_{11},\sigma_{22}$ where $\widehat{q}_\alpha(\widehat{F}_{\bPi}) \geq 0$ when using a grid search; the solid black line indicates the boundary of these pairings using numeric optimization. }
       \label{fig:aids_resilience_sets_2}
\end{figure}

\begin{figure}[ht]
    \centering
    \includegraphics[width=0.8\textwidth]{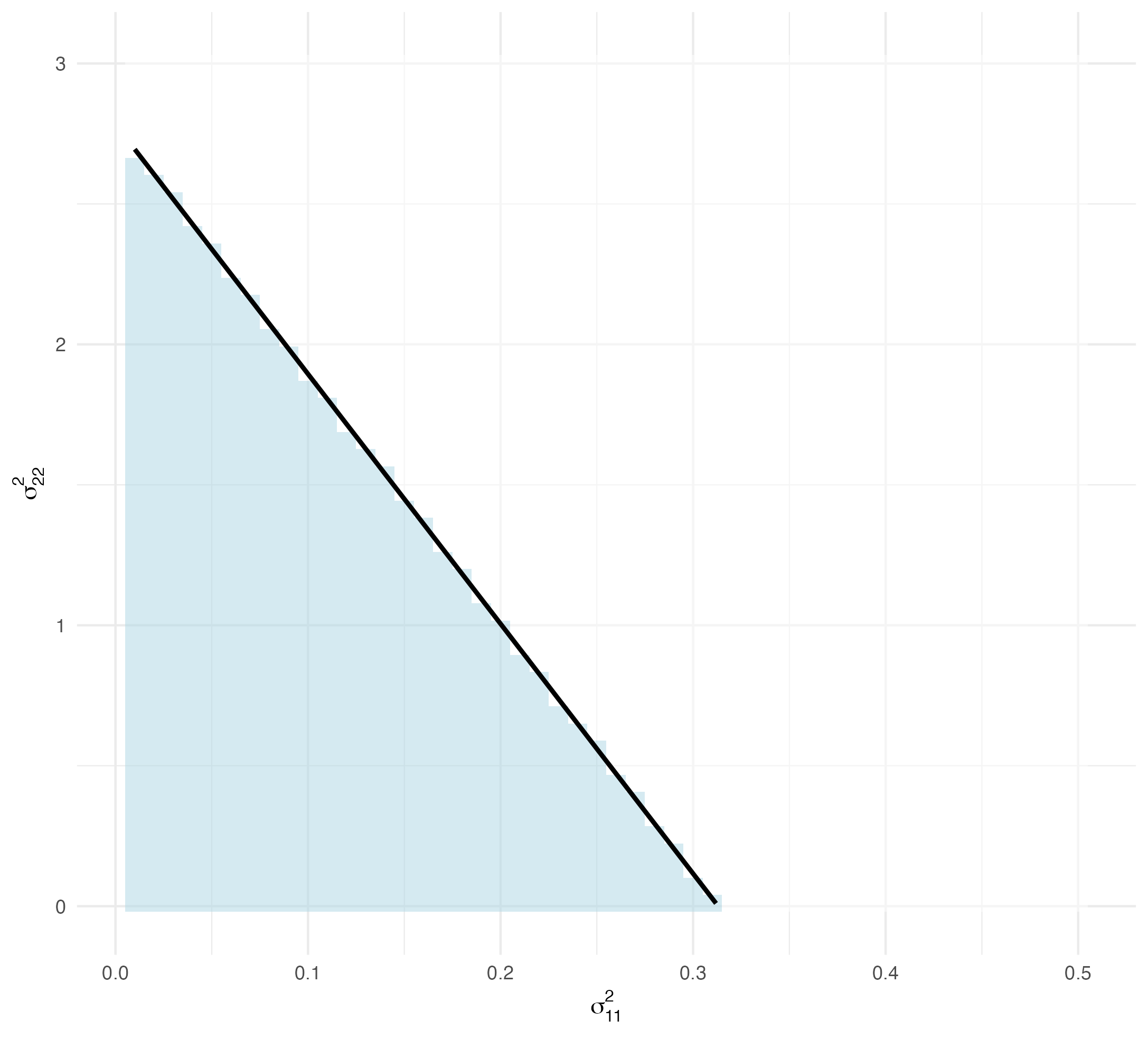}
    \caption{Estimated resilience set, $\widehat{\Omega}_q =  \{\bPi \mid \widehat{q}_\alpha(\widehat{F}_{\bPi}) \geq 0\}$, for the HIV data application using the Fourier series algorithm parameterized by $\bSigma$, with an $\alpha=0.10$; blue shading indicates the identified pairings of $\sigma_{11},\sigma_{22}$ where $\widehat{q}_\alpha(\widehat{F}_{\bPi}) \geq 0$ when using a grid search; the solid black line indicates the boundary of these pairings using numeric optimization. }
       \label{fig:aids_resilience_sets_3}
\end{figure}

In addition to using the fixed Gaussian Process algorithm in the main text, we applied our proposed approach using the polynomial and Fourier series algorithms. For both, we set  $\bSigma$ to be a diagonal matrix in the form $diag(\sigma_{11}^2, \sigma_{11}^2, \sigma_{22}^2, \sigma_{22}^2)$, and set $\sigma_{11}^2$ and $\sigma_{22}^2$ as functions of the average residuals between the kernel smoothed estimates and actual observed $Y$ values in the combined treatment and control groups. All standard errors (SE) were estimated via the bootstrap and $\alpha=0.10$. Specifically, for the Fourier series approach, we used $\sigma_{11}^2=0.409, \sigma_{22}^2=0.205$ and set the periods of the terms to be 0.5, 0.25, 0.1, and 0.05 times the range of $S$ values in Study A. This resulted in $\widehat{p}= 0.144$ with an SE of $0.027$, and $\widehat{q}_{\alpha}= -0.184$ with an SE of $0.114$. For the polynomial class algorithm, $\sigma_{11}^2=0.0818, \sigma_{22}^2=0.020$ which resulted in $\widehat{p} = 0.036$ with an SE of $0.014$ and $\widehat{q}_{\alpha} = 0.308$ with an SE of $0.106$. \ref{tab:hiv_results} summarizes the results using each algorithm. These results illustrate how the choice of functional class can meaningfully affect the estimated resilience quantities. The Fourier series algorithm, which allows for highly oscillatory deviations from the estimated conditional means, yields a negative resilience bound, suggesting a potential for the surrogate paradox under extreme or rapidly fluctuating alternatives. In contrast, the polynomial class (and Gaussian Process) constrains the shape of possible deviations more, leading to a more conservative (positive) bound. Lastly, we estimated the resilience set for both the polynomial and Fourier series approach, displayed in Figures \ref{fig:aids_resilience_sets_2} and \ref{fig:aids_resilience_sets_3}.

\clearpage

\begin{table}[ht]
\centering
\begin{tabular}{lccccc}
\hline
\text{Algorithm}  & $\widehat{p}$ & $\widehat{p}_{\text{SE}}$ & $\widehat{q}_\alpha$ & $\widehat{q}_{\alpha,\text{SE}}$ \\
\hline
Gaussian Process & 0.014 & 0.014 & 0.408 & 0.114 \\
Fourier & 0.144 & 0.027 & -0.184 & 0.114 \\
Polynomial & 0.036 & 0.014 & 0.308 & 0.106 \\
\hline
\end{tabular}
\caption{Results of different algorithms applied to the HIV data application, where $\widehat{p}$ denotes the estimated resilience probability, $\widehat{p}_{\text{SE}}$ denotes the standard error of the estimated resilience probability obtained via bootstrap, $\widehat{q}_\alpha$ denotes the resilience bound with $\alpha=0.10$, and $\widehat{q}_{\alpha,\text{SE}}$ denotes the standard error of the estimated resilience bound obtained via bootstrap,}
\label{tab:hiv_results}
\end{table}

\clearpage

\clearpage
\begin{landscape}
\begin{table}
\caption{Simulation results investigating robustness to misspecification of the basis class, showing estimates of the resilience probability, $p_0$, and resilience bound $q_{\alpha}$ for $\alpha=0.10$ in Setting 2 (Gaussian Process [GP] data generation) using the Fourier series algorithm, in Setting 4 (polynomial data generation) using the Gaussian Process algorithm, and in Setting 9 (Fourier series data generation) using the polynomial algorithm, in terms of bias, empirical standard error (ESE), and average standard error (ASE) estimated via the bootstrap. Note that for each setting, the algorithm used for estimation corresponded to the data generation (e.g. Gaussian Process algorithm was used for estimation in Settings 1-3); in Appendix C we additionally investigate robustness to misspecification of the basis class in these settings.\\}
\label{tab:misspecified-simulations}
\begin{tabular}{|c|l|l|c|c|c|c|c|}
\hline
\multicolumn{8}{|l|}{\textbf{Resilience probability, }$\mathbf{p_0}$} \\
\hline
Setting & Description & Algorithm Used & Truth & Estimate & Bias & ESE & ASE \\
\hline
2 & paradox possible, GP& Fourier & 0.383 & 0.425 & -0.043 & 0.068 & 0.067 \\ \hline
4 & paradox probable, polynomial & GP & 0.540 & 0.490 & 0.050 & 0.200 & 0.178 \\ \hline
9 & paradox unlikely, Fourier &Polynomial &  0.014 &  0.001 &  0.013 &  0.001 &  0.002 \\ \hline
\multicolumn{8}{|l|}{\textbf{Resilience bound, }$\mathbf{q_{\alpha}}$} \\
\hline
Setting & Model & Truth & Estimate & Bias & ESE & ASE \\
\hline
2 & paradox possible, GP & Fourier & -0.688 & -1.154 & 0.465 & 0.214 & 0.195 \\ \hline
4 & paradox probable, polynomial & GP & -3.323 & -0.770 & -2.554 & 0.359 & 0.348 \\ \hline
9 & paradox unlikely, Fourier  & polynomial &  0.500 & 0.680 & -0.180 & 0.053 & 0.053 \\ \hline
\end{tabular}
\end{table}
\end{landscape}

\clearpage

\clearpage
\bibliographystyle{biom}
\bibliography{Surrogate_master.bib}

\end{document}